\documentclass[aps,floatfix,multicol,epsfig,amssymb,amsmath,ifthen,preprint,letterpaper]{revtex4}
\usepackage{amsmath}
\usepackage{graphicx}

\begin{document}

\title{\bf Simulation of Si:P spin-based quantum computer architecture}

\author{Angbo Fang$^1$, Yia-Chung Chang$^1$, J.R. Tucker$^2$}
\address{$^1$Department of Physics, University of Illinois at Urbana-Champaign,\\
1110 West Green Street, Urbana, Illinois 61801}
\address{$^2$Department of Electrical and Computer Engineering\\
University of Illinois at Urbana-Champaign, Urbana, Illinois
61801}

\date{\today}
\begin{small}
\begin{abstract}
We present a systematic and realistic simulation for single and
double phosphorous donors in a silicon-based quantum computer
design. A two-valley equation is developed to describe the ground
state of phosphorous donors in strained silicon quantum well (QW),
with the central cell effect treated by a model impurity
potential. The dependence of valley splitting of the donor ground
state on QW width and donor position are calculated and a
comparison with valley splitting of the lowest QW states is
presented. The oscillation of valley splitting is observed as the
QW width or donor position is varied in atomic scale.  We find
that the increase of quantum well confinement leads to shrinking
charge distribution in all 3 dimensions. Using an unrestricted
Hartree-Fock method with Generalized Valence Bond (GVB)
single-particle wave functions, we are able to solve the
two-electron Sch${\ddot o}$dinger equation with quantum well
confinement and realistic gate potentials.  The lowest singlet and
triplet energies and their charge distributions for a neighboring
donor pair in the quantum computer(QC) architecture are obtained
at different gate voltages. The effects of QW width, gate
voltages, donor separation, and donor position are calculated and
analyzed. The gate tunability and gate fidelity are defined and
evaluated, for a typical QC design.  Estimates are obtained for
the duration of $\sqrt{SWAP}$ gate operation  and the required
accuracy in voltage control.  A strong exchange oscillation is
observed as both donors are shifted along [001] axis but with
their separation unchanged. Applying a gate potential tends to
suppress the oscillation.  The exchange oscillation as a function
of donor position along [100] axis is found to be completely
suppressed as the donor separation is decreased. The simulation
presented in this paper is of importance to the practical design
of an exchange-based silicon quantum computer.
\end{abstract}
\end{small}
\maketitle

\section{Introduction}
    Recent years there have been increased research activities
toward the understanding, characterization, and  fabrication of a
scalable quantum computer (QC) [1], due to the invention of
efficient quantum algorithm and quantum error correction. Among
more than twenty kinds of implementation proposals, the
silicon-based QC architecture has attracted much enthusiasm, ever
since Kane's seminal proposal [2] in 1998. The existing
microelectronics technology can provide significant insights and
huge resources during the development of a silicon-based QC,
giving it significant advantage over other QC candidates.

    In Kane's scheme, the nuclear spins of $ ^{31}P$ dopant atoms
embedded in a silicon host, were proposed as the elementary QC
units, or qubits.  The resonance frequency for a nuclear spin can
be tuned by the hyperfine interaction between electron and nuclear
spins, which is controlled via a gate electrode, the so-called
A-gate, over each qubit.  Thus, a globally applied a. c. magnetic
field can realize arbitrary rotations on selected nuclear spins. A
J-gate (combined with A-gate), placed over the middle of
neighboring donors, allows a controllable qubit-qubit interaction
by varying the exchange coupling between the mediating donor
electrons via the change of potential barrier between neighboring
qubits.  Obviously, this proposal requires a delicate spin
transfer between electrons and nuclei for read-in-read-out of
quantum data and multi-qubit operation.

     There also exist variations to the original
Kane theme [3,4,5,6], to circumvent the difficulties or simplify
the QC implementation. Vrijen et al [3] suggest utilizing the full
power of band-structure engineering and epitaxial
heterostructures. The spin of a bound electron in a donor atom is
used as the qubit. The qubit can be selectively tuned in and out
of resonance under gate-control to realize single-qubit operation,
by taking advantage of g-factor variations available in SiGe
heterolayers. Larger surface gate biases could be employed to
couple neighboring qubits  in a controllable way by displacing the
bound electrons deep into Ge-rich layers. However, this scheme
suffers from the complexity of g-factor engineering and possible
qubit loss due to ionization when the bound electron is dragged
into the high-g region.

 In this paper we study another modified version of Kane-type
 architecture proposed in Ref. [7].  We follow the encoding
    scheme by DiVincenzo et. al. [8], where universal quantum
    computation can be realized for composite 3-spin qubits, with
    only nearest-neighbor Heisenberg exchange interaction. A logic qubit, encoded by
    the subspace of three neighboring phosphorous donor electron spins, is
    shown in Fig. 1.  Logic zero is represented by spins 1 and 2 in
    the singlet state S, and spin 3 up. Logic one is a linear
    combination of triplet states $T_{+}$ and $T_{0}$ for spins 1
    and 2, with spin 3 down and up, respectively, to preserve the
    overall spin quantum numbers. Initializing to logic zero is
    achieved by cooling the system in a large magnetic field to
    polarize spin 3, while inducing an even greater exchange
    coupling $J_{initial}>2\mu_{B}B>> k_{B}T$, between spins 1 and
    2 to produce the equilibrium spin singlet. Typical parameters
    for $10^{-6}$ initialization error at T=100 mK are $B\sim 1 T$
    and $J_{initial}\sim 200 \mu eV$.

    There are two kinds of top gates to control qubit
    operations, shown in Fig. 1, similar to the A-gates and J-gates in Kane's
    proposal. One-qubit operations are implemented with the
    nearest-neighbor exchange in four or fewer steps, eliminating
    the need for gated ESR rotations. The two-qubit CNOT operation
    can also be performed by the nearest-neighbor exchange
    within a 1D array, in 19 steps[8].    Readout can
    be performed with a single electron transistor(SET) incorporated
    close to individual qubit, by utilizing
    spin-charge transduction[9].

 In our design (Fig.~1), phosphorous donors are placed at substitutional sites on a plane
 inside
the $Si/Si_{1-x}Ge_x$  quantum well(QW). We use the composition
$x=0.3$ for SiGe, which yields approximately a quantum well band
offset of $\Delta E_c = 300 meV$ with respect to the barriers. The
strained silicon QW not only prevents ionization of the donors
during gate operation, but also reduces the valley degeneracy of
the donor ground state.  The $Si_3N_4$ layer right below the top
metallic gates, is introduced to produce the desired gate
potential and exchange coupling.  The planar SET shown in Fig. 1
can be patterned into the P $\delta$-layer by STM along with the
individual P donor qubits, in the same lithographic step.

   Our device presented in
    Fig.1 is somehow similar to that of Friesen et. al.[4], where one uses a Si/SiGe quantum well
    to confine electrons vertically and use top gates to confine electrons laterally.
    Electron spins in such "quantum dots" act as qubits,
    with top gate voltages tuning the exchange coupling between
    qubits. However, the use of phosphorous donor electrons in our
    device provides a much smaller qubit size and a more densely
    integrated QC as we scale up the number of qubits.
   Furthermore, as we shall see in this paper,
   the available strong gate-assisted exchange coupling allows a much faster gate
    operation than that in the Friesen et. al. proposal, which would
    enable faster computation and more easily satisfy the condition
    for gate operation duration v.s. qubit decoherence time required by quantum error correction
    principle. By using naturally bound donor electrons as qubits,
    we avoid additional top gates used to provide lateral
    confinement and  the problem of precisely
    controlling the number of electrons in each quantum dot.

    The issue of decoherence of qubits is of  critical importance
    for the construction of a realistic QC. The quantum error
    correction principle has set severe limits on the amount of
    tolerable decoherence for feasible QC operations. More than
    $10^4$ reliable quantum gate operations need to be
    performed during the coherence time, while the fidelity of
    gate operation could be lost for too short a gate
    duration. The spin decoherence in
    Si:P system has been intensively studied both experimentally and theoretically
    in recent years[10,11,12,13,14].
    Due to the weakness of electron-phonon and spin-orbital interaction,
    the longitudinal spin relaxation time $T_1$ at low temperature can be as long as
    $10^3$ seconds[15], while the transverse spin dephasing time $T_2$ is of
    the order of 60 ms at 7K for isotopically purified ${}^{28}$Si:P, as detected by spin echo technique[11].
    Theoretically it is expected that $T_2$ can be as long as  $2T_1$ in isotopically purified
    ${}^{28}$Si.  This extraordinarily long decoherence time is another
    attractive aspect of spin-based Si:P QC architectures.

    Due to the attractive features of the Si:P QC, recently there have been
    much increased research activities
on Si:P donors, such as the re-estimate of spin decoherence
time[10-14], the stark effect and gate-induced ionization of
single donor[16-18], and most importantly, the exchange coupling
for a donor pair[19-24].

     The present paper is organized as follows. In the next
section we review the shallow donor problem in Si, incorporating
the details of the Si band structure.  We present the necessary
formalism for the multi-valley effective mass equation for single
donor and the unrestricted Hatree-Fock  approach with generalized
valence bond(GVB) wave functions for coupled donor pair.  We study
in Section III the energy level splitting and charge distribution
for a quantum-well-confined single donor. Section IV presents the
most important results of our work. The exchange coupling for a
donor pair is extensively studied, including the effect of quantum
well size, the oscillatory behavior v. s. donor position shift,
and gate voltage dependence. The calculation of the realistic gate
potential is presented in the appendix by solving the Poisson
equation with appropriate boundary conditions. In the last section
we summarize our work and present relevant remarks on the Si:P
based QC architecture.

\section{CHARACTERIZATION OF PHOSPHOROUS DONORS IN SILICON}

    The problem of impurities in semiconductors was extensively
    studied in the sixties and seventies of last century, along with
    the growth of the semiconductor industry [25].
The most common and successful approach to solve the single
impurity problem is to utilize the effective mass approximation
(EMA) [26]. For semiconductors like silicon, there are several
equivalent valleys in the conduction band due to crystal symmetry,
and the original hydrogenic EMA needs to be modified to
incorporate the interactions between degenerate valleys known as
''valley-orbit coupling''.  The main contribution to the
valley-orbit coupling arises from the short-range core part of the
impurity potential.  The impurity potential is usually not
Coulombic near the core, and it is difficult to calculate it from
first principles due to the complication of exchange and
correlation effects between the donor electron and core electrons.
Therefore, in the early applications of EMA, the agreement between
theories and experiments is good for excited states but poor for
ground states which have a high probability inside the core. Later
on many theoretical attempts [27] were made to account for the
discrepancies between the hydrogenic binding energies $E_H$ and
the observed donor binding energy, $E_0$. This difference, mainly
due to the greatly reduced of screening of Coulomb potential
inside the so-called "central-cell" region, is commonly referred
to as the "chemical shift" or "central-cell correction".

     Experimentally [28], for phosphorous donors in
     bulk silicon, the zero-field binding energies are
     $E_b+\Delta_0 = 45.59 meV$ (the nondegenerate ground state with A symmetry),
     $E_b+\Delta_0-\Delta_E = 33.89 meV$ (two-fold degenerate state with E symmetry), and
     $E_b+\Delta_0-\Delta_T=32.58 meV$ (three-fold degenerate state with $T_2$ symmetry)V.
     Here $E_b$ is the binding energy of six-fold degenerate ground
     states obtained by single-valley EMA.
     Kohn and Luttinger's theory [26] yield $E_b\approx 29 meV $, while the more elaborate work by
     Faulkner [29] gives $E_b\approx31.27 meV$. $\Delta_0$, $\Delta_E$,
     $\Delta_T$, known as central-cell corrections, are positive , and in contrast to $E_b$, depend on
     the particular impurity species. For Si:P donors,
     $\Delta_0\approx 16.6 meV$ according to Kohn and Luttinger (or
     14.3 meV according to Faulkner), $\Delta_E \approx 13.0 meV$
     and $\Delta_T =11.7 meV$.

     It is pointed out by Pantelides and Sah [30] that the concept of
chemical shifts or central-cell corrections are ill-defined in the
one-valley EMA framework, except for isocoric impurities (where
impurity and host have the same number of core electrons, e.g.,
Si:P donor), and the "chemical shift" must arise almost entirely
from intervalley mixing, which causes the splitting of the
six-fold degenerate ground state level of donors embedded in
silicon as observed experimentally. Therefore, a multi-valley
effective mass equation (MVEME) is needed to correctly solve the
impurity problem in a host with multi-valley conduction band.

     The first effective-mass equation for a multi-valley band was obtained by
     Twose as reported by Fritzsche [31]. For a single substitutional donor
     situated at $\bf R$ in silicon, whose band structure has six equivalent valleys,
     the electron wave function in real space can be written as:
\begin {equation}
\Phi({\bf r}-{\bf R})=\sum_{i} F_i({\bf r}-{\bf R}) e^{{\bf
k}_i\cdot({\bf r}-{\bf R})} u_i({\bf r})
\end {equation}
where $u_i({\bf r})$ is the periodic part of the Bloch functions
for silicon lattice.  $F_i$ is the envelope function for the i-th
valley, with $i=\pm x, \pm y, \pm z$ corresponding to six valleys
centered at the band minima $\pm k_0 \hat{\mathbf x}$, $\pm k_0
\hat{\mathbf y}$, and $\pm k_0 \hat{\mathbf z}$, respectively.
$k_0=0.86\times {2\pi}/{a_0}$, with $a_0=5.43 \AA$ the lattice
constant for silicon crystal. The Twose's multi-valley effective
mass equation(MVEME) reads:
 \begin {equation}
\sum_{i}\alpha_i e^{i({\bf k_i}-{\bf k_j})\cdot{\bf r}}
[T_i(-i\nabla)+ U({\bf r})-E] F_i({\bf r}) =0
\end {equation}
where the set of coefficients $\alpha_i$ is determined by crystal
symmetry group. $U({\bf r})$ is the impurity potential and $T_i
({\bf k})$ is the expansion of the silicon band structure
$E^0({\bf k})$ at the i-th minimum. For example, around the
minimum along $+x$ direction, located at $+k_0 \hat {x}$, the
energy may be expanded as
\begin {equation}
E^0({\bf k})\simeq \frac {\hbar^2}{2m_l}(k_x-k_0)^2+ \frac
{\hbar^2}{2m_t}({k_y}^2+{k_z}^2).
\end {equation}
where $m_l$ and $m_t$ are the longitudinal and transverse
effective masses, respectively.

  The MVEME [Eq.~(2)] is essentially a set of coupled equations
for the envelope functions associated with different valleys. If
the inter-valley mixing, also called valley-orbital interaction,
is negligibly small, the envelope functions for different valley
nearly coincide and the set of MVEME equations reduce to one
single-valley EME. That's the origin of the six-fold degenerate
ground state for donors in silicon obtained by Kohn and Luttinger
[26].  Only by taking into account the coupling between different
valleys we can obtain correct splitting of the ground state.  For
a substitutional donor in silicon, the impurity potential has
tetrahedral symmetry. Valley-orbit coupling mixes the six Bloch
functions belonging to different valleys with three kinds of
symmetries according to the irreducible representations of the
$T_d$ group. The linear combinations of them  form a singlet with
$A_1$ symmetry, a doublet with $E$ symmetry, and a triplet with
$T_2$ symmetry.  The appropriate linear combinations are
\[A_1:(X+\bar{X}+Y+\bar{Y}+Z+\bar{Z})/\sqrt{6} ;\]
\[E: (X+\bar{X}-Y-\bar{Y})/2,{}(2Z+2\bar{Z}-X-\bar{X}-Y-\bar{Y})/\sqrt{12} ;\]
\begin{equation}
T_2:
(X-\bar{X})/\sqrt{2},{}(Y-\bar{Y})/\sqrt{2},{}(Z-\bar{Z})/\sqrt{2}.
\end{equation}
Here  $X$, $\bar{X}$, $Y$, $\bar{Y}$,$Z$ and $\bar{Z}$ are the
labels for the six Bloch functions, according to the directions of
conduction band minima.

   Eq.~(2) can be solved once we know the form of the impurity
   potential $U({\bf r})$.  For substitutional donors, such as phosphorous donors in silicon,
the impurity potential, $U({\bf r})$, consists of two parts:
$U_b({\bf r})$ and $U_s({\bf r})$, where $U_b({\bf r})$ is
essentially the difference between the potential of an impurity
ion and a host ion, and $U_s({\bf r})$ arises from the
redistribution of valence electrons in the crystal, caused by the
presence of $U_b({\bf r})$. For phosphorous donors in silicon, the
point-charge model is reliable and linear response theory can be
validly applied to solve the impurity potential[27,30]. The
resulting impurity potential in k space can be written as

\begin{equation}
 U({\bf q})= U_b({\bf q})/\epsilon({\bf q})
\end{equation}
with $U_b({\bf q})$ given by the Fourier transform of
\begin{equation}
U_b({\bf r}) = -n e^2/r +W_b({\bf r})
\end{equation}
where the first term is the Coulomb potential of one point charge
and $W_b({\bf r})$ denotes the short-range contribution. The
q-dependent dielectric screening function, instead of simply the
constant $\epsilon_0$ in the hydrogenic impurity potential, is
used to take care of the correct screening effect both near and
far away from the core. A set of exponential functions may be used
to simulate the dielectric screening function[30].

For Si:P donors, $W_b$ is very small and localized within about
one lattice constant.  Similar to the short-range behavior of the
dielectric screening function, it's difficult to calculate.
Therefore we include all of the short range effects in the
modelled dielectric screening function.  For convenience we will
approximate the dielectric screening effect by a set of isotropic
Gaussian functions.  It is found that, a single-parameter Gaussian
function is adequate for simulating the (inverse of) dielectric
screening function calculated in Ref. [32] within several lattice
constants ($a_0=5.43 \AA$) and yield the correct long-range limit.
Explicitly, the dielectric function used in this paper has the
following form:
\begin{equation}
\frac {1}{\epsilon(r)}=\frac{1}{\epsilon_0}
[1+(\epsilon_0-1)\exp(-\alpha_c r^2)]
\end{equation}
where $\alpha_c$ is treated as an empirical parameter. We find
that $\alpha_c = 1.13 a.u.$ gives reasonable agreement with the
dielectric function in the range of more than two lattice
constants and it yields the correct intervalley splitting of the
Si:P ground state.

   Eqs. (1), (2), (3) and (7) constitute all we need to adequately describe the electron wave
   function and energy levels for an isolated Si:P donor.
   For our specific QC architecture, phosphorous donors are
   confined in a Si/SiGe quantum well.  The large in-plane strain
   presented in the quantum well further reduces the crystal
   symmetry.  As discussed in
   Refs. [21] and [33] within perturbation theory, the interplay of
   strain and valley-orbit effect makes only two valleys relevant for the low-lying states.
   For the quantum well shown in Fig. 1,
   the minimums of two lowest energy valleys are at $k_x =k_y=0$
   and $k_z =\pm k_0$. Their energies are well separated from those at the minimums of
   the other four higher energy valleys by more than 100 meV. Thus,
   only the +z and -z valleys contribute to the ground state.
   Explicitly, we are left with two kinds of linear combinations of
   Bloch wave functions,
\[T^{+}_{z}: symmetric: (Z+\bar{Z})/\sqrt{2} ;\]
\begin{equation}
T^{-}_{z}: antisymmetric: (Z-\bar{Z})/\sqrt{2}.
\end{equation}
The symmetric $T^{+}_{z}$ state is the ground state for a single
donor
   at the center of a strained quantum well.

   Analogous to Eq.~(1), for a single donor located at ${\bf R}=0$,
   the two-valley wave function can be written as
\begin {equation}
 \Phi({\bf r})=F_{+z}({\bf r}) e^{+k_0\cdot z}u_{+z}({\bf r})+
 F_{-z}({\bf r}) e^{-k_0\cdot z}u_{-z}({\bf r})
\end {equation}

   The corresponding two-valley effective mass equation with external field is given by
\[ [T_{+z}(-i\nabla)+ U({\bf r})+V_{ext}-E] F_{+z}({\bf r})+\]\[e^{+i 2
k_0 z} [T_{-z}(-i\nabla)+ U({\bf r})+V_{ext}-E] F_{-z}({\bf r})=0
\]\[
 e^{-2 i k_0 z} [T_{+z}(-i\nabla)+ U({\bf r})+V_{ext}-E]
F_{+z}({\bf r}) \]\begin{equation}+ [T_{-z}(-i\nabla)+ U({\bf
r})+V_{ext}-E] F_{-z}({\bf r})= 0
\end{equation}
where $T_{+z}=T_{-z}\equiv T_z$ has an operator form as follows:
\begin{equation}
T_z(-i\nabla) = - \frac {\hbar^2}{2m_l} \frac{\partial^2}{\partial
z^2}- \frac {\hbar^2}{2m_t} (\frac{\partial^2}{\partial
x^2}+\frac{\partial^2}{\partial y^2}).
\end {equation}

  The envelope functions for the two valleys, $F_{+z}({\bf r})$ and
  $F_{-z}({\bf r})$, are coupled together, as shown in Eq.~(10). We do not assume the equivalence
  of $F_{+z}({\bf r})$ and $F_{-z}({\bf r})$ as, e.g.,  in Refs. [4] and
  [21], since we shall consider the situation where the system
 lacks reflection symmetry about $z$.

  To solve the electron wave functions and energies, we need to
choose some well-behaved basis functions to expand the two
envelope functions.  We modify the set of bases used in our
previous paper[20] (in which we study Si:P donors within
spherically averaged EMA) to take into account  the anisotropy of
the silicon band structure and the broken symmetry brought by the
external field.

  We use a set of  two-dimensional(2D) anisotropic Gaussian
  functions $\{ \beta_j(y,z) \}$ to describe the freedoms
  in y and z directions. Two different kinds of Gaussian functions,$\{ \exp [-\alpha_i
  (y^2+z^2/\xi^2)]; i=1,...,n_e \}$ and $\{ z \exp [-\alpha_i
  (y^2+z^2/\xi^2)]; i=n_e+1,...,n_r=n_e+n_o \}$, are used to simulate
  the envelope functions with even and odd
  symmetry (with respect to z). The optimum value for the anisotropy factor $\xi=0.57$ is
  obtained by minimizing the ground state energy.  We note that both
  even and odd symmetric bases are necessary when the external
  potential lacks reflection symmetry with respect to the doping
  plane(z=0).  The set of Gaussian
  parameters $\{\alpha_i\}$ are optimized such that a linear
  combination of the 3D Gaussian functions $\{\exp(-\alpha_i r^2) \}$
  best resembles the 1s wave function of a hydrogen atom, while a linear
  combination of $\{z \exp(-\alpha_i r^2) \}$ best resembles the $2p_z$
  orbital.

   In our multi-qubit QC architecture, a line of substitutional donors are
arranged along the [100] axis, and neighboring qubits can be
coupled together mainly by the wave-function overlaps along the x
axis. To facilitate our calculation and to better describe the
inter-donor coupling, we place the system in an artificial 1D box
with infinite potential barriers. The center of the box is at
$x=0$ and the box size $L$ is chosen to be large enough (at least
10 nm from the donor ion to either side wall) so that it has
negligible effect on the donor binding energy and charge
distribution. With the introduction of the box, we can use a set
of orthogonal sine functions, $\{
\sqrt{2/L}\sin[\frac{m\pi}{L}(x+L/2)], m=1,...,n_x \}$, to
describe the electron freedom along the x direction. Again, such a
set of functions includes both odd (even m) and even (odd m)
functions of $x$.  This is needed when the donor ion is shifted
from the box center (which is the case when more than one donor
are present) and/or a nonsymmetric external potential is
introduced. Combining the set of sine functions and $\{
\beta_j(y,z), j=1,...,n_r \}$, we get a set of bases:
\begin{equation}
B_{jm} =\sqrt {2/L}\sin [\frac {m\pi}{L}(x+L/2)] \beta_j(y,z)
\end{equation}
where $jm=(j-1)\times n_x+m$ and $N=n_r \times n_x$ is the total
number of our basis functions.

  The suitability of this construction of bases was discussed and
  verified in a previous publication [20] by comparing our numerical results to exact
  results in various limits.

  Now we can expand the two valley-dependent envelope functions as a linear
  combination of our bases. We write:
\[
F_{+z}=\sum_{jm=1}^{N} c_{jm} B_{jm}\]  \\
\begin{equation}
F_{-z}=\sum_{jm=1}^{N} c_{jm+N}B_{jm}.
\end{equation}

  Substituting Eqn. (13) to Eqn. (10),
we obtain a matrix equation for the 2N-dimensional vector $\{c_j,
j=1,...,2N \}$:
\begin{equation}
 \sum _{n=1}^{2N}H_{mn}c_n = E \sum _{n=1}^{2N}O_{mn}c_n
\end{equation}
where $\{H_{mn} \}$ and $\{O_{mn} \}$ are  $2N \times 2N$
Hermitian matrixes, which can be written as $2 \times 2$ block
forms in terms of four $N \times N$ matrixes:
\begin{equation}
 H_{2N \times 2N} = \left[ \begin{array}{cc}
                             H^{11} & H^{12} \\
                             H^{21} & H^{22}
                             \end{array}  \right]
\end{equation}
where $H^{11}$ and $H^{22}$  are the intra-valley Hamiltonian
matrixes for +z and -z valley, respectively, while $H^{12}$ and
$H^{21}$ describe the inter-valley coupling. Obviously,
$H^{21}{}^{*}=H^{12} \equiv H^{DV}$, since $H_{2N \times 2N}$ is
Hermitian. We also have $H^{11}=H^{22}\equiv H^{SV} $  since we
have chosen the same set of basis functions to expand the envelope
functions for the two valleys. The decomposition of $O_{2N \times
2N}$ has similar properties ($O^{21}{}^{*}=O^{12} \equiv O^{DV}$
and $O^{11}=O^{22}\equiv O^{SV} $). Explicitly, the matrix
elements of $H^{SV}$ and $H^{DV}$ are written as
\[ H^{SV}_{ij} =\int d{\bf r} B_i({\bf r}) [T_{z}(-i\nabla)+
U({\bf r})+V_{ext}]B_j ({\bf r});\]
\begin{equation}
H^{SD}_{ij} = \int d{\bf r} B_i({\bf r}) [T_{z}(-i\nabla)+ U({\bf
r})+V_{ext}]B_j ({\bf r}) e^{i2k_0z}.
\end{equation}
where it is understood that the operators inside the square
brackets only act on the envelope basis functions but not on the
valley phase factor. $O^{SV}$ and $O^{DV}$ can be obtained by
replacing $[\cdots]$ in Eq.~(16) by 1.

  Our main equation, Eq.~(14), is a generalized eigenvalue equation composed of  a $2N\times 2N$
  hamiltonian matrix H and a $2N\times 2N$ overlap matrix. Since the sine functions in $x$ are already orthogonal, we
  only need to perform the Gram-Schmidt
  orthogonalization process on the set of $y,z$ dependent gaussian
  functions to convert the generalized eigenvalue equation into a
  standard eigenvalue equation.  We have
\begin{equation}
{\tilde H}^s {\tilde C} = E {\tilde C}
\end{equation}
where ${\tilde H}^s$ is a $2N \times 2N$ Hermitian matrix and
${\tilde C}$ denote a 2N-dimensional eigenvector. The full single
donor electron wave function can be constructed from ${\tilde C}$
according to Eqs.~(9) and ~(13), once the Bloch functions $u_{\pm
z}({\bf r})$ are known.

 To study a donor pair coupled by mutual Coulomb interaction,
  we need to build the appropriate form of the two-electron wave
  function and the effective two-electron Hamiltonian(Born-Oppenheimer approximation is implicitly assumed).

  Like the potential term in the one-electron MVEME, the electron-electron(e-e) interaction term
  should also include contributions from different valleys.
 However, for well-separated donors, the short-range effect on the
 mutual Coulomb interaction is negligible. Therefore,
the e-e interaction takes a simple form,
\begin{equation}
U_{ee}({\bf r_1}, {\bf r_2})= \frac {1}{\epsilon_s|{\bf r_1}-{\bf
r_2}|}
\end{equation}
where $\epsilon_s\equiv \epsilon(0)=11.4$ is the static dielectric
constant.

  The two-electron Hamiltonian can thus be written as
\begin{equation}
H({\bf r_1}, {\bf r_2})= {\tilde H}^s({\bf r_1})+{\tilde H}^s({\bf
r_2})+ U_{ee}({\bf r_1}, {\bf r_2}).
\end{equation}

  For a well-separated donor pair, the two-electron spatial
  wave function with the lowest energies can be constructed from
  the one-electron orbits of both donors, according to the Heitler-London approximation.
  The lowest two-electron state is a singlet in the absence of  magnetic
  field and is a symmetric combination of one-electron wave-function
  products, while the second lowest two-electron state is a
  triplet and antisymmetric with respect to the interchange of
  spatial indexes of the two electrons.  Explicitly, we write
\begin{equation}
\Psi_{\pm}({\bf r_1},{\bf r_2}) = \frac 1 {\sqrt{2(1+S^2)}} [
\Phi_L({\bf r_1})\Phi_R({\bf r_2})
 \pm \Phi_R({\bf r_1})\Phi_L({\bf r_2})],
\end{equation}
where the subscript +(-) denotes the singlet (triplet) state,
$\Phi_L$($\Phi_R$) denotes a one-particle wave function localized
at the left (right) donor site, and $S\equiv|\langle\Phi_L| \Phi_R
\rangle|$ the wave function overlap.

   Our goal is to solve the Schr\"{o}dinger equation
\begin{equation}
H({\bf r_1},{\bf r_2})\Psi_{\pm}({\bf r_1},{\bf r_2})=E_{\pm}
\Psi_{\pm}({\bf r_1},{\bf r_2}).
\end{equation}

   The Heisenberg exchange coupling strength, J, which plays an essential role for
   a lot of solid-state QC proposals, is given by the energy difference
   of the lowest singlet and triplet states: $J=E_{-}-E_{+}$.

   As we did successfully before [20], we use the unrestricted
   Hartree-Fock method to solve the lowest singlet and triplet states in the form
of Eq.~(20).  Let the expansion coefficients in terms of the
orthogonalized bases $\{ \tilde B_n({\bf r}); n=1,...,2N \}$ for
$\Phi_R$ and $\Phi_L$ be $R_n$ and $L_n$, respectively. Assume
that in a given iteration, we have already known the expansion
coefficients $R_n$ for $\Phi_R({\bf r})$.  The above two-particle
eigen-value equation can be reduced to a single-particle
eigen-value problem by projecting it into the state $\Phi_R$. The
projected eigen-value equation within the basis now reads
\[
\sum_n [\langle n'|H^s|n \rangle  +\langle \Phi_R|H^s|\phi_R
\rangle \delta_{n',n}
 \pm \langle n'|H^s|\Phi_R \rangle R_n
\]
\[ \pm R_{n'} \langle \Phi_R|H^s|n \rangle
+ \langle n',\Phi_R|U_{ee}|n,\Phi_R \rangle \pm \langle
n',\Phi_R|U_{ee}|\Phi_R,n \rangle  ] L_n \]
\begin{equation}
=E \sum_n (\delta_{n',n} \pm R_{n'} R_n ) L_{n}.
\end{equation}
Thus, $\Phi_L({\bf r})$ can be solved via the standard
diagonalization procedure within the one-particle basis.  The
newly obtained $\Phi_L({\bf r})$ is then used to solve the new
$\Phi_R({\bf r})$, and we do it iteratively until the singlet and
triplet ground state energies and wave functions converge.

  As we discussed and examined in our previous paper [20], the
  above method dealing with a two-particle problem, which we call UHF with generalized
  valence bond wave functions, does a better job than the usual
  single-determinant HF method, or the HL
  approximation. The floating-phase Heitler-London approach, proposed by Koiller
 et. al. [23], corresponds to a special case of our method
 where a single parameter describing the relative phase of  $\Phi_L$ and  $\Phi_R$
 is allowed to vary to improve the naive application of the usual HL approximation.
Compared to the configuration interaction(CI) method, which uses a
set of  two-particle wave function bases(each basis corresponding
to a Slater determinant) to do an exact diagonization, our
proposed method uses 2 relaxed determinants to expand the wave
function.  However, we adopt our method instead of the CI method
because it can handle the well-separated two-donor problem
adequately and requires much less computational resources, while
the large number of bases needed in the CI method makes it
currently unpractical.

  For convenience we define the effective Bohr radius
\begin{equation}
{a_B}^* = \frac{\epsilon_s\hbar^2}{m^*e^2}
       = 0.529 \AA \frac{\epsilon_s}{m^*/m_0}
       \simeq 23.4 \AA
\end{equation}
and the effective Rydberg (donor binding energy)
\begin{equation}
Ry^*= \frac{e^4 m^*}{2\hbar^2 \epsilon_s^{2}} = 13.59 eV
 \frac{m^*/m_0} {\epsilon_s^{2}}\simeq 27 meV.
\end{equation}
where $m_0$ is the free electron mass, $m^*$ is the spherically
averaged effective mass, $m^*=(\frac 1 3 {m^*_l}^{-1} + \frac 2 3
{m^*_t}^{-1})^{-1} =0.258 m_0$ with $m_l=0.911 m_0$ and $m_t=0.190
m_0$ for silicon.

Throughout the paper, if not specified, we will use the effective
atomic units (a.u.) in which distance is measured in ${a_B}^*$ and
energy measured in $Ry^*$.

\section{Single Phosphorous Donor in Silicon Quantum Well}
    In this section we will study the single Si:P donor confined in a Si/SiGe quantum well
    with the method we developed in Section II.  Although there
    have been enormous publications in the past on the theory of isolated
    donors in silicon [27], only in recent years have silicon donors
    been studied in a general, inhomogeneous environment due to its
relevance to gate control of the bounded electrons in quantum
computing context. The present authors [20] studied the effects of
a uniform electric field and a one-dimensional parabolic external
potential on a pair of Si:P donors, within one-valley spherical
EMA. Martins et. al. [34] addressed the behavior of shallow donors
in silicon under a uniform electric field, using a tight-binding
model.  Smit et. al. analyzed in Ref. [16] the effect of a small
nearby gate on a hydrogenlike impurity in a semiconductor up to
field ionization and in Ref. [18] the small electric field
dependence of the lowest energy states in donors and acceptors,
beyond the single-vally EMA, by applying symmetry arguments and
perturbation theory. Blom et.al. [33] used a one-valley hydrogenic
effective-mass model with central-cell corrections for calculating
the properties of shallow donors inside or outside heterostructure
quantum wells. Very recently, Friesen [17] developed a
multi-valley effective mass theory to study the Stark effect of
substitutional donors in silicon, but with valley-orbit coupling
treated perturbatively.

   For shallow donors in a stressed Si/SiGe quantum well, the
lattice-constant mismatch lifts the degeneracy of the silicon
conduction band and only the two lowest-energy (+z and -z) valleys
are populated. The strain-induced splitting causes the same amount
of energy shift for the left two valleys, so we assume its effect
is completely included in the band offset $U_s$.

    The external field potential in Eq.~(10) is thus composed of
two parts: the square QW potential $V_{QW}$ and the gate
potential. Details for calculating the gate potential profile can
be found in Appendix.

     The quantum well potential has the following form:
\begin{equation}
  V_{QW}(z) = \left \{ \begin{array}{ll}
         U_s & \mbox{if $|z-z_m|> W/2$}\\
         0 & \mbox{otherwise}
         \end{array}
          \right.
\end{equation}
where $W$ is the width of the quantum well and $z_m$ the position
of the QW central plane.

     First we study the case where only $V_{QW}$ is present (without the gate potential).
We solved the two-valley EME for a single donor placed in the
middle of the well, with the QW width varied from 4 nm to 16 nm.
The single donor atom is also at the center of the artificial box.
To study the lowest energy states, we only need to consider bases
with even symmetry with respect to both x and z. Ten Gaussian
functions and 120 plane wave functions are used to construct the
singe-particle bases. The single parameter in the dielectric
function, $\alpha_c$, if taken as 1.13 a.u., yields a intervalley
splitting of 3.2 meV between the ground state and the first
excited-state, when the QW width is as large as 16 nm. Note that
the ground state energy in wide well limit is approximately -34.21
meV, close to the experimental value of the 1s states with $T_2$
symmetry for a phosphorous donor in bulk silicon. The value of
intervalley splitting for a single P donor in a strained Si
quantum well, although not available experimentally, was estimated
to be 3.3 meV in large strain limit by Koilter et. al. [21] within
perturbation theory, while Blom et. al. [33] estimated the
central-cell shifts by relating it to the bulk shifts with
symmetry consideration. The shift of the lowest two states from
the QW one-valley effective-mass value is given by [33]:
\begin{equation}
\frac {|\Psi_{QW}({\bf r}_0)|^2}{|\Psi_{1s}({\bf r}_0)|^2}
[-\Delta_0 +\left\{ \begin{array}{l}
                   2\Delta_E/3 \\
                   \Delta_T \end{array} \right\}].
\end{equation}
where ${| \Psi_{QW}({\bf r}_{0})|}^2$ and ${| \Psi_{1s}({\bf
r}_0)|}^2$ are the QW and bulk envelope function amplitudes near
the impurity center, respectively.  The values of $\Delta_0$,
$\Delta_E$, and $\Delta_T$ are given in Section II, for
phosphorous donors in bulk silicon.

   In the large QW width limit, the QW-bulk ratio for envelope function
amplitudes should be of the order of unity, which gives a valley
splitting of about 3 meV, consistent with our calculation. Figure
2 shows our calculation of the Si:P ground-state and
first-excited-state energies and their splitting as a function of
QW width in large scale.  We see that both the ground and excited
state energies increase as we decrease the QW width, as a
consequence of larger kinetic energy for stronger confinement. The
intervalley splitting also increases in a similar way.  With
Eq.~(26), Blom et. al. predicted the same behavior for the valley
splitting as the QW width is varied.  They argued that, the ratio
in Eq.~(26) is expected to increase as the QW is narrowed, due to
additional confinement, therefore, the intervalley splitting
should also increase.

   This can be verified by inspecting the charge distribution.
We plot in Fig 3 (a) and (b) the 1D averaged charge distribution
(obtained by integrating out the other two coordinates in the 3D
charge density) along z-axis and x-axis, respectively. The inset
of Fig. 3(b) shows the 1D averaged charge distribution along
y-axis. First we note that, the curves of $\rho_x$ and $\rho_y$
are very similar to each other, although $\rho_x$ is slightly more
spread out, due to different basis functions we use for $x$ and
$y$ directions (by symmetry they should coincide with each other).
The charge distribution along z axis oscillates with a period
$2\pi/\Delta k\equiv\pi/k_0\approx 3 \AA$, resulting from the full
wave function construction in Eq.~(9). Due to influence of QW
barriers, the charge distribution is more concentrated around the
donor nucleus for narrower confinement, as can be seen from the
comparison of $\rho_z$ for the QW width at 6nm and 10nm.

To our surprise, the charge distribution in the direction parallel
to the QW plane, shown in Fig.3 (b), also shrinks toward the
impurity center as the QW is narrowed. This behavior is
counterintuitive since we would have expected the shrinking of
wave function in one direction would accompany dispersing in other
directions. The synchronization of charge redistribution in all
the 3 dimensions upon the change of 1D confinement,  is due to the
cooperative efforts to maintain the 3D shape of 1s orbital so as
to lower the ground state energy. We expect that the QW
confinement should also have a significant effect on the exchange
coupling of two neighboring donor electrons inside the quantum
well, as we shall discuss in Sec. IV, since the dispersion of
charge distribution would enhance the electron wave function
overlap as the QW confinement is reduced.

   We plot in Fig. 4(a) the valley splitting $\Delta E_P$ of the Si:P ground
   state as the QW width varies in small scale. The calculated
   points cover QW width range from about 8nm to 10nm,
   with a spacing of $a_0/4\simeq 1.36\AA$.  It
   clearly shows an oscillatory behavior and  the
   oscillation magnitude decays as the QW width increases, while the
   oscillation period is approximately one lattice constant ($a_0$).

   For comparison, we also calculate the valley splitting of the lowest QW states, $\Delta E_W$.
Such a valley splitting has been re-examined recently for Si/SiGe
heterostructures within a 2-band tight-binding model [35,36].
 Here it is done by solving the two-valley EME without the impurity
 potential.  The boundary matching conditions at the QW interfaces are
 neglected, since the lowest QW states are mostly confined inside the
 silicon quantum well for a QW width more than 8nm and a barrier height of 300 meV.
 We use the products of 20 (2D)Gaussian functions and 100 sine
functions to expand the 3D wave functions of the lowest QW states.
Fig 4(b) shows the QW valley splitting for the QW width varing
from 50 to 80, in units of $a_0/4$. First we note that the
magnitude of $\Delta E_W$ is  approximately several tenths of meV
for the QW width range shown, consistent with the tight-binding
calculation of Ref. [36]. Compared to the valley splitting of
impurity states $\Delta E_P$, $\Delta E_W$
 is about 6 times smaller.  This significant difference between the magnitude of $\Delta E_W$
 and $\Delta E_P$ is due to the presence of central-cell correction brought by
 the impurity potential.  Second, similar to $\Delta E_P$,  $\Delta
 E_W$ also oscillates and decays as the QW width increases.
 Comparing  the pattern of Fig. 4 (b) with Fig. 4 (a), we see that the oscillation
 of $\Delta E_W$ is much more frequent than $\Delta E_P$ on this
 smallest scale. However, if we view the oscillations at a larger scale (as in Ref. [36]), say,
 in units of $a_0/2$, $\Delta E_W$ will oscillate less frequent than $\Delta E_P$, as shown by the dotted lines
 in Fig. 4 (a) and (b).  Note that the oscillation period on the
 $a_0/2$ scale is about $3.5 a_0$, nearly the same as the oscillation shown in Fig. 3 of
 Ref. [36].

   Compared to the sophisticated microscopic tight-binding
 calculation, we conclude that, our simple calculation, although
 it has overlooked the delicate boundary matching condition for
 finite-height barriers, is able to produce similar
 results. The oscillatory behavior shown in Fig. 4 (a) and (b), is
 due to the presence of multiple propagating states in the well,
 which is inherently included in our two-valley effective mass
 approximation, as shown in the composition of the full wave function in
 Eq.~(9).

   We also study the intervalley splitting for the Si:P
   donor states as the donor atom is moved away from the center plane of
   the quantum well (but still kept in the center of the artificial box).
   In this case, the reflection symmetry of the
   QW potential is broken and we need to include both even and
   odd basis functions with respect to $z$.
   We use 10 Gaussian functions in the form of $\{ \exp [-\alpha_i
  (y^2+z^2/\xi^2)]\}$(with $\alpha_i$ optimized to resemble the 1s
  orbital of the hydrogen atom [37]) and  6 Gaussian functions in the form of
  $\{ z \exp [-\beta_j (y^2+z^2/\xi^2)] \}$ (with $\beta_j$ optimized to resemble the
  $2p_z$ orbital of the hydrogen atom [37])  to take care of the
  mixed symmetry. Eighty sine functions with even parity
  is used for the description of freedoms along x axis.

    We plot in the inset of Fig. 4(a) the valley splitting as a function of donor
    position inside the silicon quantum well with the QW width
    fixed at 9 nm. The valley splitting also shows an oscillatory
    behavior as the donor position is shifted. The magnitude of
    oscillation increases as the donor is moved away from the
    QW center, while the oscillation shows a double-period structure,
    with periods $\sim 0.5 a_0$ and $\sim 0.75 a_0$.
  The increase of the valley splitting as the donor
is moved away from the QW center  can be explained with the help
of Eq.~(26). The QW envelope function amplitude ${| \Psi_{QW}({\bf
r}_{0})|}^2$ is expected to increase as one of the barrier is more
easily sensed by the donor bound electron (similar to the case
when the quantum well is narrowed).

   The effect of gate potential on the single donor will not be
presented here, since we focus in this paper on the gate control
of exchange coupling for a donor pair, as we shall address in the
next section. It is worth mentioning that although the spectrum
narrowing is observed (the intervalley splitting decreases as the
gate-induced electric field pulls the donor electron away from the
central cell), the ground state energy is lowered, in contrast to
the result obtained in Ref. [17] for donors in bulk silicon  where
the Stark effect of phosphorous donor in bulk silicon was studied
with a perturbative multi-valley effective mass theory.

\section{Exchange Coupling for Si:P Donor Pair}

    In this section we present systematic studies on a phosphorous
donor pair embedded in a strained silicon quantum well. The
Heisenberg exchange interaction, defined as the splitting of the
lowest singlet and triplet states, plays a central role in most
solid state QC architectures. As a result, exchange coupling has
been frequently revisited for a pair of localized electrons ever
since the first quantum dot QC proposal. Heitler-London
approximation is most frequently used, where single-particle
orbitals are first variationally determined and two-particle wave
function is constructed as in Eq.~(20). This kind of
approximation, although does fairly well for well separated
impurities (or quantum dots), it does not provide enough
flexibility to handle the strong-coupling case where electron
correlation effect becomes important.  As a better alternative, we
suggested in Ref. [20] to apply the UHF method with GVB wave
function to solve the two-electron system. In this paper we follow
this method, combining it with the MVEME, as formulated in Section
II.

  We consider a pair of coupled donors, labelled by 1 and 2 in Fig. 1.
The gate potential to control the exchange coupling is tunable via
two gate voltages,  $V_c\equiv V_{12}$ and $V_g\equiv V_{01}
\equiv V_{23}$.   The gate electrodes right above donor atoms,
$V_1$ and $V_2$ (known as A-gates) are disabled in this paper,
since they are not necessary for tuning the exchange coupling for
a donor pair (but they will be needed for the cooperation of 3
neighboring donors). The modelling of the electrostatic potential
distribution due to applied voltages on the gates is presented in
the appendix. The potential distribution in the QW region is give
by Eq.~(A9). Fig. 5 (a)-(d) shows several gate potential profiles
for different values of $V_c$ and $V_g$. We see that these two
gate voltages are indeed capable of tuning the barrier height
between two P donors in the QW.

    The width of the artificial 1D box is set to be the same as
the size of the 2-donor unit cell, $L_x = 4p$, for the donor array
with a spacing of $2p$. Since the electrons are bound to the
donors mainly within 1 ${a_B}^*$ and dispersed a little in the
presence of gate potential, the box boundaries will not affect the
charge distribution when p is larger than $4 {a_B}^*$. The box
boundaries pass through the centers of gate electrode $V_{01}$ on
the left and $V_{23}$ on the right, respectively. $V_{01}$ and
$V_{23}$ (cooperating with other gate electrodes) are typically
set to produce potential barriers high enough to suppress the
coupling between the specific donor pair (1 and 2) with
neighboring donors (labelled 0 and 3, respectively) during each
step of gate operations. In this sense, our 1D box is no longer
artificial.  The central J-gate, $V_{12}$, is right above the
middle point (x=0) of the two donors.

     First we calculate the exchange coupling when the gate potential
is absent ($V_c =V_g =0$).  The donor separation, is first set to
$R\equiv 2p=10 {a_B}^*$, a generally accepted value for Si:P
spin-based QC architecture and within the access of current
nano-fabrication technology[38]. We vary the QW width from 4nm to
16nm and plot the corresponding exchange splitting in Fig. 6. In
the wide-well limit, the exchange splitting converges to about
$1.4\times 10^{-6} Ry^*$, on the same order of magnitude but a
little smaller, compared to the exchange coupling for two
hydrogenic impurities in bulk silicon estimated via the asymptotic
formulae[39]. However, it is about 40 percent larger than the
numerically calculated value by single-valley hydrogenic effective
mass approximation in Ref. [20].  Our current two-valley
calculation is more reliable, since we take into account the
effective-mass anisotropy, central-cell correction to the impurity
potential, and inter-valley coupling.

    We see clearly from Fig. 6 that the exchange splitting J depends
strongly on the QW width.  It decreases as the quantum well is
narrowed. The exchange splitting at $W_d=6nm$ is about one order
smaller than that at $W_d=16nm$.  Comparing their single-particle
overlaps, $S \equiv |\langle \Phi_L| \Phi_R \rangle|$ for singlet
states (the overlaps are always zero for triplets according to our
construction of two-particle wave functions) with each other, we
have $S=5.8 \times 10^{-4}$ for $W_d=6nm$ and $S=1.7\times
10^{-3}$ for $W_d=16nm$, respectively.  As we narrow the quantum
well from 6nm to 4nm (slightly less than 2 ${a_B}^*$), the
exchange splitting is reduced further by one order of magnitude
and the overlap $S$ decreases to $2.0\times 10^{-4}$.   This is
just what we expected, since from Sec. III (Fig. 3) we know that
the charge distribution is shrinking toward the impurity centers
in all three dimensions with the increasing of QW confinement.
This confinement-induced shrinking effect decreases the overlap
between neighboring donor electrons, and therefore, reduces their
exchange coupling.

    The inset of  Fig. 6 illustrates the effect of QW-width
variation on small scale on the exchange splitting, where the QW
width is varied from 58 to 74 monolayers (approximately from 8 to
10 nm) and the spacing of width variation is one monolayer (1
monolayer=$a_0/4 \simeq 1.36 \AA$).  Different from the
oscillatory behavior of the valley splitting for single donor
ground state shown in Fig. 4(a), the exchange splitting varies
with the QW width monotonously even at the atomic scale.

     Now we switch on the gate voltages
$V_g$ and $V_c$.  The QW width is fixed at 10nm, the appropriate
value for producing the desired range of the exchange coupling for
our QC architecture. We shall fix the value of $V_g$ and vary
$V_c$ (the central gate voltage), since the potential barrier
height between the coupled donors is mainly influenced by voltage
difference of the central gate and side gates.  Fig. 7(a) shows
the influence of the central gate voltage ($V_c$) on the exchange
coupling for donors separated by 10 ${a_B}^*$, with $V_g$ fixed at
-2.0, -1.8 and -1.4 V, respectively.   We notice that voltage
dependence of the exchange splitting can be approximately fitted
by an exponential curve (shown as straight line in logarithm-scale
plot), similar to our previous simulation where single-valley
hydrogenic EMA is employed and the gate potential is modelled by a
one-parameter one-dimensional parabolic function. The three lines
corresponding to three different values of side gate voltages, are
nearly parallel to each other, which means that the logarithmic
slope is independent of the gate voltages and it only depends on
the intrinsic geometry settings of the QC architecture, such as
the donor separation, QW width, and distance of doping plane from
the top gates. Therefore, we may define the logarithmic slope as a
measure of the gate voltage tunability (with respect to the
exchange coupling) as
\begin{equation}
\eta = \frac {\partial \log{J}}{\partial V},
\end{equation}
where we have omitted the subscript of $V_c$ and obviously, $\eta$
has the dimension of $V^{-1}$.   Note that in Ref. [40], $\partial
J/\partial V$ is similarly defined as the susceptibility of a
device to voltage errors.  Here, our gate voltage tunability is
better defined since it is constant throughout the appropriate
working voltage range and it characterizes the intrinsic device
properties.

  Since $\eta$ is independent of $V_g$, we may focus on the case with side gate voltages
fixed at $V_g = -1.8 V$.  By tuning the central gate voltage
$V_c$, we are able to vary the exchange coupling by more than
eight orders of magnitude. The full range is not shown in Fig.
7(a), but we can further reduce the exchange splitting to as low
as $10^{-11} Ry^*$ by lowering $V_c$ to -3.0 V for $V_g = -1.8 V$.
The exchange splitting at $V_g=-2.1 V$ approximately equals the
value of $\approx 10^{-6} Ry*$ in the absence of gate potential
($V_c =V_g=0$). Reliable quantum computation requires a much
smaller J-coupling than the gate-off value for qubit isolation. In
Ref. [7], it is argued that the 'off-state' coupling will need to
be $J_{off}\approx 10^{-12} eV$ or less. We see that the
J-coupling at $V_c=-3.0 V$ can fulfill this requirement.

  Furthermore, the large initialization requirement $J_{on}\approx 200 \mu eV$
(required to initialize the logic 0 state of the 3-donor qubit)[7]
is also within our tunability.  At $V_c=-0.6(-0.8)$, we can reach
a J-coupling of approximately $400 (93) \mu eV$.

  However, as discussed in Ref. [20], we can not arbitrarily
enhance the exchange coupling by lowering the potential barrier
between donors.  To illustrate this, we plot in Fig. 8 the 1D
averaged charge distribution along x axis for the lowest singlet
and triplet states at different gate voltages. Each plot shows a
two-peak structure with peaks near the donor nuclei. Except for
Fig.~8(a) and (d) where the coupling and overlap is weak and the
charge distributions for singlet and triplet almost coincide,
singlet state has a larger probability than the corresponding
triplet state in the central region between donors, due to the
Pauli exclusion principle[20]. In Fig.~9(a), although the gate-off
exchange coupling is nearly equal to that at $V_c = -2.1V$ and
$V_g =-1.8V$, their charge distributions are different. For the
gate-off case, the two charge distribution peaks are symmetrically
centered at the donor sites, while for the case of $V_c = -2.1V$
and $V_g =-1.8V$, they are more spread out and become nonsymmetric
with respect to the donor locations. However, the averaged donor
electron distances for these two cases, are nearly the same, which
is the reason that they have close magnitudes in exchange
coupling. Fig.~8(d) shows the charge distribution for the
'off-state' configuration.  The peak centers are shifted
significantly away from each other with a negligible overlap, and
are more dispersed than those in Fig.~8(a).  For the case of $V_c
=-0.8$ shown in Fig.~8(b), the singlet is only slightly different
from the triplet, with slightly larger (smaller) weight in the
central (donor site) region, and the peak centers remain close to
the donor locations as in the case of configurations of (a),
although they are pulled toward each other to give an enhanced
overlap. In Fig.~8(c), when the (attractive) gate potential
becomes even stronger, the difference between singlet and triplet
charge distributions becomes more significant, and the weights in
the central region are comparable to those in the donor regions.
For this case, the gate potential is strong compared to the
impurity potential, and electrons have a large probability to be
trapped in the central region. This situation is undesirable for
QC implementation, since it may complicate the spin dynamics for
gate operations, and more importantly, produce extra decoherence
channels and reduce gate fidelity. The $V_c=-0.8$ configuration is
relatively much safer and can nearly produce the desired exchange
coupling in the large limit.

   In Fig.~7(b), we compare the gate voltage dependence of
J-coupling for a pair of donors separated by R=8, 10, and 12
${a_B}^*$, respectively.  First, we notice that for R=8 and 12
${a_B}^*$, both the corresponding curves deviate (but not
significantly) from straight lines on logarithmic scale.  The
three lines are crossed with each other near $V_c =-0.7V$.  The
logarithmic slopes, indicting the gate voltage tunability on the
exchange coupling, are remarkably different. As the central gate
voltage is tuned from -2.2 to -0.6 V, the variation range of
exchange coupling is 2, 5, and 9 orders, for R=8, 10 and $12
{a_B}^*$, correspondingly. Obviously, larger donor separation
corresponds to large gate tunability on exchange coupling.
However, as we will see below, larger gate tunability also leads
to more stringent requirement for voltage control accuracy.

    With a time-dependent voltage pulse, the exchange coupling
which controls the spin dynamics  varies accordingly as the time
progresses, and spin states can be manipulated in a controllable
way.  As a prototype for gate operation, we consider the adiabatic
$\sqrt{SWAP}$ operation for two neighboring donor electrons,
implemented with a voltage pulse $V_s(t)$.  In principle, the
pulse shape of $V_s(t)$ can be arbitrary, although the accumulated
phase by the exchange coupling must satisfy the following
relation:

\begin{equation}
\int_{\tau_s}dt J(V(t))/\hbar = \pi/2.
\end{equation}
Here $\tau_s$ is the gate operation time and the exchange coupling
is dependent analytically on the central gate voltage via

\begin{equation}
J(t)\simeq \alpha \times \exp(\eta V(t)),
\end{equation}
where $\alpha = 0.224 eV$ and $\eta = 6.9 V^{-1}$ for $R=10
{a_B}^*$ and $V_g\equiv -1.8  V$.

   To get a realistic estimate of the gate operation time, we assume a symmetric linear
(central gate) voltage pulse, defined by
\begin{equation}
V(t)= \left \{ \begin{array}{ll}
   V_0+2\Delta V (t/\tau_s+0.5) &\mbox{if $-\tau_s/2\leq t\leq 0$;} \\
   V_0-2\Delta V (t/\tau_s-0.5) &\mbox{if  $0\leq t \leq \tau_s/2$}
                \end{array}
       \right.
\end{equation}
where the voltages during the whole gate operation span a range of
$[V_0, V_0+\Delta V]$. We have already shown that [-3.0,-0.8] is
an appropriate voltage range for $R=10{a_B}^*$, which can give a
sufficiently small 'off-state' exchange coupling and  large
'on-state' exchange coupling without disturbing the two-electron
charge distribution significantly. The maximum exchange coupling
is achieved right at the middle of gate operation, corresponding
to $V_0+\Delta V$.

   Plugging Eqs.~(29) and (30) into Eq.~(28) with $V_0 =-3.0V$ and $\Delta V =2.2V$,
the gate operation time is found to be $t_s \simeq 0.2 ns$.
Compared to the experimental transverse decoherence time of
$T_2\approx 60 ms$[11], the proposed pulse gives a relative error
of $\sim 10^{-7}$,  well satisfying the fault tolerant computation
requirement of a relative error less than $10^{-4}$. On the other
hand, the gate operation time is restricted from below by the
adiabatic gating requirement[41],
\begin{equation}
\tau_s > \tau_{min}:= max\{\hbar/\delta \epsilon_1, \hbar/\delta
\epsilon_2\},
\end{equation}
where $\delta \epsilon_1$ and $\delta \epsilon_2$ are the
single-particle lowest level spacing and two-particle lowest
singlet level spacing, respectively.  $\delta \epsilon_2$ is
roughly determined by the difference of the on-site and inter-site
Coulomb interaction. As discussed in Ref.[41], for ultra small
quantum dots, $\delta \epsilon_1$ is usually much larger than
$\delta \epsilon_2$. However, this is not the case for
multi-valleyed semiconductors such as Si. The presence of gate
potential can cause single-particle spectrum narrowing as
discussed in the last section. $\delta \epsilon_1$, given by the
intervalley splitting of the single donor ground state, is
approximately $3.0 meV$ when gate potential is absent, while with
gate potentials at $V_c=-0.8V$ and $V_g=-1.8V$ we have $\delta
\epsilon_1 \sim 0.43 meV$. In comparison, $\delta \epsilon_2$ is
$\sim 0.78 meV$ at the same voltages.  We find that when $V_g$ is
fixed at $-1.8V$ and $V_g$ varies in the voltage range
[-3.0,-0.8], $V_c=-0.8V$ corresponds to the smallest value for
both $\delta \epsilon_1$ and $\delta \epsilon_2$. Therefore,
according to Eq.~(31), we get an estimate of the gate operation
time, $\tau_s > 1.5 ps$ on the adiabatic lower restriction. This
requirement is also well satisfied by our proposed linear voltage
pulse with a gate operation time of 0.2 ns.

This gate-induced narrowing of single-particle level spacing
requires an operating temperature to be about six times lower than
the gate-off estimate($\sim 30K$), to avoid unwanted thermal
transition. To improve the operating temperature, we may increase
the contribution of the other four valleys by reducing the Ge
composition in the SiGe barrier (i.e. reducing the in-plane
strain), decrease the quantum well width, or re-design the gate
potential profile. On the other hand, the increase of
single-particle level spacing or valley splitting, requiring the
single-particle wave functions to be more localized around the
impurity centers, would unavoidably reduce the available maximum
exchange coupling between neighboring donor electrons(if the donor
separation keeps unchanged). Thus the temperature requirement
further complicates the issue for optimal design of the QC
architecture.  However, this problem may be circumvented by the
recently proposed multilevel encoding scheme, where a logic qubit
is encoded by a subspace of multiple physical levels. The full
quantum-computational fidelity is maintained in the presence of
mixing and decoherence within logical subspaces. For details of
multilevel encoding, see Ref. [42].

    The gate voltages that control exchange couplings between
neighboring donor electrons cannot be produced at arbitrary
accuracy. This results in deviation from the desired pulse shape
and exchange coupling, causing gate errors.  The instantaneous
relative error rate can be quantified by $\delta J(t)/J(t)$ and
the accumulated error rate $1-F =\int_{\tau_s} \delta J(t) dt/(\pi
\hbar/2)$ indicates the loss of gate operation fidelity.

    Due to the exponential dependence of exchange coupling on the
    gate voltages, the instant error rate can be simply related to the gate tunability
    and voltage fluctuation by
\begin{equation}
\delta J(t)/J(t) = \eta \delta V(t).
\end{equation}

   With this relationship we can obtain a lower bound of the
gate fidelity:
\begin{equation}
F_{-} = 1-\eta \delta V_0,
\end{equation}
where $\delta V_0$ is the maximum magnitude of voltage uncertainty
or fluctuation during the switching time, limited from below by
the available control electronics.   The other parameter affecting
the gate fidelity, is the gate tunability $\eta$ intrinsic to the
QC architecture.  We note that, to reduce the loss of fidelity,
$\eta$ should not be too large.  This restriction, combining with
the requirement of enough tunability so as to achieve the desired
'initialization' and 'off-state' exchange coupling, calls for an
appropriate parameter zone (such as the donor spacing, the
distance of doping plane to the surface gates, and the width of
the silicon quantum well) for the realistic quantum computer
device design.

  For the typical $\eta \sim 7$ for two donors separated by 10 ${a_B}^*$,
a voltage control accuracy of at least $1.4 \times 10^{-5} V$ is
required to meet the error threshold (1 accumulated error in
$10^4$ operations) of fault tolerant computation.  This voltage
accuracy is about the same order as in Ref. [4], where the authors
estimated it with a flattop exchange coupling pulse for the SiGe
quantum dot QC architecture.

  Koilter et. al. [19] analyzed the effect of donor position shift on
exchange splitting for a well-separated donor pair in bulk
silicon, with the Kohn-Luttinger envelope wave function and within
Heiter-London approximation. They pointed out the importance of
precise donor positioning in QC implementation, due to the
exchange oscillation caused by inter-valley interference inherent
in silicon band structure. Later on, a few authors confirmed this
kind of exchange oscillation under the perturbations due to
applied strain [21] or gate fields [22], still within standard HL
framework.  Recently, a reconfirmation has been done with a
floating-phase Heitler-London approach [23].   A similar approach
was used for studying Si/SiGe quantum dots in Ref. [43]. In that
paper, the authors demonstrated that, with strong confinement
potential and large in-plane strain, there is no atomic-level
oscillations in the single-electron wave function, and
consequently no exchange oscillation, in contrast to donors in
bulk silicon.

 The phosphorous donors in our QC architecture are embedded in a
 strained Si
 quantum well with no additional lateral confinement, thus lying between
 the two extreme cases: donor electron in bulk silicon and electron confined in
 gated Si/SiGe quantum dots. Furthermore, as we pointed out in Sec II, our
approach to the two-localized-electron problem is more advanced
than the Heitler-London approximation or its modified version. Now
we examine whether the donor-position dependent exchange
oscillation exists with and without the presence of gate potential
(and with four valleys suppressed by the in-plane strain).

  First we study the case where two donors are shifted along z axis from the QW
  central plane by the same amount with their separation
  remaining unchanged.  The calculating points are spaced by
  $a_0/4$, the smallest atomic spacing along z axis. It is worth
  mentioning that, in Ref. [19], donor positions are shifted in
  units of $10\%$ of the nearest-neighbor distance in silicon
  lattice,  which is nonphysical for substitutional phosphorous
  donors.

  In Fig.~9(a), we plot the exchange coupling  versus shift of donor position (up to 3$a_0$)
  along the [001] direction. The system is symmetric with respect to the QW central
plane in the absence of gate potential, and both donors remain on
the [100] axis with their separation fixed at 10${a_B}^*$.
 We see from Fig. 9(a) that, the exchange coupling does
oscillate as both donors are shifted along z axis, contrary to the
monotonous behavior shown in the inset of Fig. 6, where the QW
width is varied on atomic scale. The oscillation magnitude is
largest as the donors are shifted from the QW center by a single
unit ($a_0/4$) and it decays as donors are moved away from the
center plane.  The upper-bound of the $\Delta J/J$ for the
smallest displacement, is about $5\sim6 \%$. As can be seen from
the trend of Fig. 9(a),  the exchange coupling tends to increase
as both donors are shifted away.  The inset shows the
corresponding singlet overlaps, similar to the pattern of exchange
coupling.

  We plot in Fig. 9(b) the exchange coupling as a function of
  donor position shift when a gate potential is switched on.  We
  find that, the exchange oscillation still exists, but the
  over-all exponentially growing behavior is preserved.

  Now we study the exchange coupling as both donors are shifted
  along their coupling direction, i. e., [100] axis.  The smallest possible donor position
  shift (and donor separation change) is $a_0/2$, to keep both donors in the QW central plane.
  To better describe the delicate change of the wave function overlap upon position shift along
  $x$ axis, we use 280 since function(for x-direction freedom and 3
Gaussian functions (for $y$ and $z$) to construct the
single-particle
  basis.  The parameters in the 3 Gaussians, are adjusted to take
  care of the central cell correction so that correct valley splitting is reproduced for the single donor
  ground state. The 1D box width is set to be $25 {a_B}^* (\sim 60 nm)$, to further
  reduce possible artificial effects of box boundaries.  We plot
  in Fig. 10 the exchange splitting  for donor
  separation varied from 70 to 90 $a_0/2$ (or 190 to 244 $\AA$).
  The figure is plotted in logarithmic scale, as the exchange splitting has an
  exponential dependence on donor
  separation. The dotted line is an exponential fitting curve.
  We find that the fitting works fairly well. Compared to the exchange oscillation in Fig. 9(a),
  the exchange splitting is almost monotonically increasing as the donor
  separation is reduced. Only for separation larger than $85
  a_0/2$,  exchange splitting is no longer monotonic and residual oscillation is
  present.  This behavior can be qualitatively understood in the
  following way. The valley-interference induced exchange
  oscillation is analogous to noise, which is prominent when
  the exchange coupling (analogous to signal) is weak.  The exponential dependence of exchange
  splitting on donor separation, tends to suppress the oscillation
  for small enough separation (strong enough exchange coupling).
  If we take the exponential fitting curve as a reference, the deviation
oscillates strongly from both sides of the reference curve at
large separation and decays as separation is decreased.

  Based on our calculation, it is worth to make a few remarks on the importance of precise donor
  positioning to fault tolerant quantum computation.  First, the
  deviation of donors from any host atom site is highly
  undesirable, since the electronic properties of an interstitial
  donor is remarkably different from a substitutional donor.
  Second, a recent report shows that the STM-based lithographic technology
 can position single P atoms into Si with $\sim 1$ nm accuracy[44].
  If unfortunately,a donor atom is shifted away from the desired site,
  it will result in large computation error (due to the deviation of exchange coupling
from desired value) and qubit measurement error (due to the qubit
leakage to undesired subspace). However, this should be able to be
detected in the calibration process described in Ref. [7], and we
can either adjust the gate voltage to reduce error or reject the
erroneous qubit.

\section{SUMMARY AND CONCLUDING REMARKS}

  In this paper we have presented a systematic study on coupled
  phosphorous donors in a strained silicon quantum well,
  relevant for silicon-based quantum computer design.  Following the
  multi-valley effective mass approach, we have developed a
  two-valley equation for the single donor.  By turning the
  valley-coupled Sch$\ddot{o}$dinger equation into a single matrix
  eigenvalue problem, we are able to solve the donor electron ground
  state energies and wave functions. The central cell correction is consistently incorporated
  by our model potential, yielding the correct valley
  splitting.

  We have studied the effects of quantum-well confinement and donor
  position shift on the valley splitting and charge distribution for the single donor
  ground state.  We find that, as the quantum well is narrowed,
  the electron wave function shrinks and larger valley splitting
  is obtained.  For a QW at fixed width, the valley splitting also increases as the donor
  approaches the edge of the QW.  We have examined the
  effects of  QW width and donor
  position variations at the atomic scale. The oscillation of valley splitting for donor electron
  ground state is observed for
  both cases.  The valley splitting of the lowest QW states is
  also calculated as a function of QW width and oscillation behavior is confirmed. We have analyzed
  these results and compared them to relevant papers by other authors.

  The unrestricted Hartree-Fock method with GVB wave functions is
  formulated and applied to a pair of coupled donors, without losing the delicate
  nature of single-particle wave functions.  The advantages of this method
  over other approaches are discussed. Accurate wave functions and energies of the
  lowest singlet and triplet
  states can be obtained. The Heisenberg exchange coupling is calculated at different QW
  widths and we find that a wider quantum well gives rise to larger
  exchange splitting. The QW width of 10 nm is appropriate for our
  proposed QC design.

  We have calculated the gate potential profile numerically by solving Poisson
  equation with appropriate boundary conditions. The effect of
  gate voltages on exchange coupling between neighboring donors is analyzed, for
  several donor separations. We find the appropriate separation and voltage
  zones suitable for qubit initialization and gate operation.  A
  gate tunability is defined to characterize the efficiency of gates
  to alter the exchange coupling.  With a linear time-dependent voltage pulse shape,
  we have evaluated the appropriate gate operation time of 0.2 ns for a $\sqrt{SWAP}$
  operation, considering both the requirements of fault
  tolerant quantum computation and adiabatic operation.  An analytic formulae for the lower
  bound of the gate
  operation fidelity is obtained, by relating it to the gate tunability and
  voltage control accuracy.  For a typical gate tunability, and to
  realize fault tolerant quantum computation,  a voltage control accuracy of
  $1.4 \times 10^{-5} V$ is necessary.

  The valley degeneracy of the Si:P donor electron ground state could
lead to oscillation of the exchange coupling as donor positions
are shifted.   With our approach to the two-electron problem, we
have calculated the exchange coupling as a function of donor
position shift perpendicular to QW plane.  Oscillation is observed
when gate potential is switched off, and switching-on of gate
potential can suppress the exchange oscillation.  The exchange
coupling as a function of donor position shift along donor
coupling axis is also calculated.  We find that the exchange
oscillation is greatly suppressed at small donor separation and
residual oscillation can be observed at large separation.  The
oscillation behaviors can be explained by analogy to the
signal-to-noise ratio.

  In conclusion, we have studied in a systematic way the ground state of
single Si:P donors and the exchange coupling of a coupled donor
pair placed in a strained silicon quantum well.  Our calculation
is of importance in evaluating  the prospects and providing
practical guidance for a P-donor silicon-based quantum computer.

{\bf Acknowledgments}

This work was supported by DARPA DAAD19-01-1-0324 and University
of Illinois Research Board.

\mbox{}

\renewcommand{\theequation}{A\arabic{equation}}
\setcounter{equation}{0}  
\section*{APPENDIX}  
 In this appendix we present calculation of gate potential profile
for the quantum computer design shown in Fig. 1.

 We start from the profile with a single gate electrode in each period.
We want to solve the partial differential equation (PDE)
\begin{equation}
 [ -\nabla^2 +V(x,z) ] \phi(x,z) =0
\end{equation}
with the following boundary conditions (BC's):

\[ \phi(x,L)=\phi_0  \; \mbox{ for all } x. \]
\begin{equation}
 V(x,z)=\left\{ \begin{array}{ll} U_0 \gg 1 \; \mbox{  for } \;
|x|<d/2 \; \mbox{  and }  \; 0<z<t, \\ 0 \; \mbox{otherwise.}
\end{array} \right.
\end{equation}

\[ \phi(x,-b)=\phi_1 \;  \mbox{ for all } x. \]
The system has reflection symmetry about $x=0$ and is periodic in
the $x$ direction with period $p$.
We divide the system into three regions:\\
(i) $t<z<L$ (the region above gate electrodes), we have
\begin{eqnarray}
 \phi_I(x,z)=&\sum_n \cos(k_n x) D_n
[e^{-k_n(z-t)}-e^{k_n(z+t-2L)} ]\nonumber\\
+&\phi_0+D_0(z-L)
\end{eqnarray}
with $k_n=2\pi n/p; n=0,1,2\cdots$.\\

 (ii) $0<z<t$ (the region with gate electrodes inside), we have
\begin{equation}
 \phi_{II}(x,z)=\sum_m \beta_m(x) [B_m e^{-q_mz} +C_m
e^{q_m(z-t)} ]
\end{equation}
where the basis functions $\beta_m(x)$ must satisfy the 1D
Schr\"{o}dinger equation,
\begin{equation}
[-\partial_x^2 + V(x)]\beta_m(x) = q_m^2 \beta_m(x).
\end{equation}
Writing $\beta_m(x)$ in the form
\begin{equation}
 \beta_m(x) = \sum_n F_{nm} S_n \cos(k_n x); \; S_n =\frac 1
{\sqrt p}[\sqrt 2 (1-\delta_{n0})+\delta_{n0}].
\end{equation}
then $F_{nm}$ can be obtained numerically by solving the
eigen-value problem
\begin{equation}
 \sum_{n'} [ k_n^2 \delta_{nn'} + V_{nn'} ] F_{n'm} = q_m^2
F_{nm}.
\end{equation}
 where
\[ V_{nn'}=U_0 S_n S_{n'} \int_{-d/2}^{d/2} \cos(k_n x)\cos(k_{n'}
x)dx = \] \begin{equation} U_0S_n
S_{n'}[\frac{\sin[(k_{n'}+k_n)d/2]}{(k_{n'}+k_n)} +
\frac{\sin[(k_{n'} -k_n)d/2]}{(k_{n'} -k_n)}].
\end{equation}

(iii) $-b<z<0$ (the region below gate electrodes), we have
\begin{equation}
 \phi_{III}(x,z)=\sum_n \cos(k_n x) A_n [e^{k_nz}-
e^{-k_n(z+2b)}]+A_0(z+b)+\phi_1.
\end{equation}
The coefficients are determined by the following BC's:\\
(i) At $z=t$, we have

\[ S_n^2 \int_{-p/2}^{p/2} \cos(k_n x) \phi_{II}(x,t) dx \]
\[= S_n^2
\int_{-p/2}^{p/2} \cos(k_n x) \phi_I(x,t) dx \]
\begin{eqnarray}
 \Rightarrow & S_n \sum_m F_{nm} [B_m e^{-q_mt} +C_m ]
 = D_n (1-e^{-2k_n(L-t)})\nonumber \\ & + \delta_{n0}[\phi_0+D_0(t-L)].
\end{eqnarray}

 Use $\epsilon_{I}{\bf E}_{I}\cdot \hat z =\epsilon_{II} {\bf E}_{II} \cdot \hat z$, we have
\[ \epsilon_{II}S_n^2 \int_{-p/2}^{p/2} \cos(k_n x)\partial_z \phi_{II}(x,t)
dx\]\[ = \epsilon_{I}S_n^2  \int_{-p/2}^{p/2} \cos(k_n
x)\partial_z \phi_I(x,t) dx
\]
\begin{eqnarray}
 \Rightarrow \eta S_n \sum_m F_{nm}q_m[-B_m e^{-q_mt} +C_m ]\nonumber \\
  = - D_n (1+e^{-2k_n(L-t)}) k_n + D_0\delta_{n0}.
\end{eqnarray}
where $\eta \equiv \epsilon_{II}/\epsilon_{I}$. Eliminate $D_n$
(including $D_0$) by combining the above two equations yields
\begin{eqnarray}
& S_n \sum_m F_{nm} \{[1+e^{-2k_n(L-t)}] [B_m e^{-q_mt} +C_m ]\nonumber \\
& +q_m[(1-e^{-2k_n(L-t)})\eta/k_n][-B_m e^{-q_mt} +C_m
]\}\nonumber
\\&  = 2\phi_0\delta_{n,0}.
\end{eqnarray}
where it is understood that $[(1-e^{-2k_n(L-t)})\eta/k_n]$ is
replaced
by $2\eta(L-t)$ for $n=0$.\\
(ii) At $z=0$, we have
\[ S_n^2  \int_{-p/2}^{p/2}  \cos(k_n x) \phi_{II}(x,0) dx \]\[
= S_n^2  \int_{-p/2}^{p/2} \cos(k_n x) \phi_{III}(x,0) dx
 \]
\begin{eqnarray}
\Rightarrow S_n \sum_m F_{nm} [B_m  +C_m e^{-q_mt}] \nonumber\\ =
A_n (1-e^{-2k_n b}) +\delta_{n0} [A_0 b+\phi_1].
\end{eqnarray}
\[ \epsilon_{II}S_n^2  \int_{-p/2}^{p/2}  \cos(k_n x) \partial_z \phi_{II}(x,0) dx \]
\[ =
\epsilon_{III}S_n^2  \int_{-p/2}^{p/2} \cos(k_n x) \partial_z
\phi_{III}(x,0) dx
\]
\begin{eqnarray}
\Rightarrow \zeta S_n \sum_m F_{nm} q_m[-B_m  +C_m e^{-q_mt}]\nonumber \\
  = A_n (1+e^{-2k_nb}) k_n +A_0\delta_{n0}.
\end{eqnarray}

  where $\zeta \equiv \epsilon_{II}/\epsilon_{III}$. Eliminate $A_n$
(including $A_0$) by combining the above two equations yields
\[
 S_n \sum_m F_{nm} \{(1+e^{-2k_nb}) [B_m  +C_m e^{-q_mt}]\]
\begin{equation} -q_m[(1-e^{-2k_nb})\zeta/k_n][-B_m +C_m e^{-q_mt}]\} =
2\phi_1\delta_{n,0}.
\end{equation}
where it is understood that $[(1-e^{-2k_nb})\zeta/k_n]$ is
replaced by $2\zeta b$ for $n=0$. Eqs. (A12) and (A15) constitute
a set of coupled linear equations for the coefficients $B_m$ and
$C_m$, which can be solved numerically.

Now we double the period of
the system ($-2p<x<2p$) and raise the voltage on the strip with
$|x|<d/2$ (within the first period) by $V_c$ and the strip with
$2p-d/2<|x|<2p$ by  $V_g$, and similarly for all other periods.
The solution to $\phi(x,z)$ in regions I and III are written in
the same form as the case with $V_g=0$, except that $k_n's$ are
replaced by $k'_n \pi n/2p; n=0,1,2\cdots$ and the unknown
coefficients are denoted $A'_n$ and $D'_n$. In region II, we need
a set of basis functions
\begin{equation}
 \beta'_m(x) = \sum_n F'_{nm} S'_n \cos(k'_n x); \; S'_n =\frac 1 {2\sqrt {p}}[\sqrt 2
 (1-\delta_{n0})+\delta_{n0}].
\end{equation}
where $F'_{nm}$ are solved similarly as in the $V_g=0$ case with
\[V_{nn'}=U_0 S'_n S'_{n'}2 [\int_{0}^{d/2} \cos(k'_n
x)\cos(k'_{n'} x)\]\[ + \int_{p-d/2}^{p+d/2} \cos(k'_n
x)\cos(k'_{n'} x)\]\[+\int_{2p-d/2}^{2p} \cos(k'_n x)\cos(k'_{n'}
x)]dx
\]
\begin{eqnarray}
=&
U_0\delta_{n,n'}+U_0S_nS_{n'}\{\frac{\sin[(k'_{n'}+k'_n)d/2]}{(k'_{n'}+k'_n)}
+ \frac{\sin[(k'_{n'} -k'_n)d/2]}{(k'_{n'} -k'_n)}\nonumber \\ &-
\frac{\sin[(k'_{n'}+k'_n)(p-d/2)]}{(k'_{n'}+k'_n)}
-\frac{\sin[(k'_{n'} -k'_n)(p-d/2)]}{(k'_{n'} -k'_n)} \nonumber
\\ & + \frac{\sin[(k'_{n'}+k'_n)(p+d/2)]}{(k'_{n'}+k'_n)}
+\frac{\sin[(k'_{n'} -k'_n)(p+d/2)]}{(k'_{n'} -k'_n)}\nonumber
\\&- \frac{\sin[(k'_{n'}+k'_n)(2p-d/2)]}{(k'_{n'}+k'_n)}
-\frac{\sin[(k'_{n'} -k'_n)(2p-d/2)]}{(k'_{n'} -k'_n)}\}.
\end{eqnarray}

 Furthermore,
we add a particular solution defined as
\begin{eqnarray}
 \phi_p(x) = \left\{
\begin{array}{lllll} V_c \; \mbox{ for } \; |x|<d/2 \nonumber \\ V_c\frac
{p-d/2-|x|}{p-d} \; \mbox{ for } \;
d/2<|x|<p-d/2\\ 0 \; \mbox{ for } p-d/2<|x|<p+d/2\\
V_g\frac {|x|-p-d/2}{p-d}  \; \mbox{ for } \; p+d/2<|x|<2p-d/2\\
V_g  \; \mbox{ for } \; 2p-d/2<|x|<2p.
\end{array} \right.
\end{eqnarray}
Note that $\phi_p(x)$ satisfies the homogeneous PDE and it leads
to correct potential difference between the adjacent metallic
strips in region II. Thus, the potential function in region II
takes the form
\[ \phi_{II}(x,z)=\phi_p(x)+ \]
\begin{equation} \sum_m \beta'_m(x)
[B'_m e^{-m\pi z/2p} +C'_m e^{m\pi (z-t)/2p} ].
\end{equation}
 After matching the boundary conditions,
we obtain two sets of coupled equations similar to Eqs. (A12) and
(A15), except that we subtract a term $[1+e^{-2k'_n(L-t)}]F_{pn}$
and $[1+e^{-2k'_nb}]F_{pn}$ from the right hand sides of these
equations.
\[F_{pn}\equiv {S'_n}^2 \int_{-2p}^{2p}
\cos(k'_nx)\phi_p(x) dx \]\[=2 V_c
{S'_n}^2\{\sin[k'_{n}(p-d/2)]\]\[+\frac d
{2(p-d)}[\sin[k'_{n}(p-d/2)]-\sin(k'_nd/2)]\]\[ -\frac 1
{p-d}[x\sin k'_nx+\cos k'_nx/k'_n]|_{d/2}^{p-d/2} \}/k'_n
\]\[+V_g\delta_{n0}- 2 V_g {S'_n}^2\{\sin[k'_{n}(p+d/2)]\]
\[+\frac {2p-d/2} {(p-d)}[\sin[k'_{n}(2p-d/2)]-\sin(k'_n(p+d/2)]
\]\begin{equation}
-\frac 1 {p-d}[x\sin k'_nx+\cos k'_nx/k'_n]|_{p+d/2}^{2p-d/2}
\}/k'_n.
\end{equation}
The region $-a<z<0$ is now filled with a dielectric (Si$_3$N$_4$)
with dielectric constant $\epsilon_a$(the dielectric screening
constants for the silicon layer and $SiGe$ layer are close to each
other, so we ignore their differences and use $\epsilon\simeq12$).
In this region, the potential function is replaced by
\[
 \phi_a(x,z)=\sum_n \cos(k'_n x)
[\bar A_n e^{k'_nz}- R_n e^{-k'_n(z+2a)}] \]
\begin{equation} +\bar A_0(z+a)+\phi_a.
\end{equation}
Matching $\phi_a(x,z)$ with $\phi_{III}(x,z)$ at boundary $z=-a$
gives
\begin{eqnarray}
 \bar A_n = [\xi(1+\gamma)+(1-\gamma)]A'_n/2 \equiv \chi^+
A'_n, \nonumber \\ R_n = [\xi(1+\gamma)-(1-\gamma)]A'_n/2 \equiv
\chi^- A'_n,
\end{eqnarray}
 and
\begin{equation}
\phi_a=A'_0(b-a)+\phi_1, \; \bar A_0 = \xi A'_0.
\end{equation}
 where
$\gamma=e^{-2k'_n(b-a)}$ and $\xi=\epsilon_{III}/\epsilon_a$.
Matching $\phi_a(x,z)$ with $\phi_{II}(x,z)$ at boundary $z=0$
gives
\begin{eqnarray}
& S'_n \sum_m F'_{nm} \{(\chi^+ + \chi^- e^{-2k'_na}) [B'_m  +C'_m
 e^{-q'_mt}]\nonumber\\
&-q'_m[(\chi^+ - \chi^- e^{-2k'_na})\zeta/k'_n][-B'_m +C'_m
e^{-q'_mt}]\} \nonumber \\ & = 2\phi_a\delta_{n,0}-[\chi^+ +
\chi^- e^{-2k'_na}]F_{pn}.
\end{eqnarray}
where $\zeta=\epsilon_{II}/\epsilon_a$. It is understood that
$(\chi^+ - \chi^- e^{-2k'_na})\zeta/k'_n$ will be replaced by
$2\zeta [b+(\xi-1)a]$ for $n=0$.
 \mbox{}

\mbox{}


\newpage

\begin{figure}
\caption{ The composite 3-spin "universal exchange" qubit of Si:P
donors in a Kane-type architechture with the integrated SET
readout.  From bottom to top, the heterostructure cross section
consists of a thick, n-doped ground layer, a 28nm undoped
$Si_{0.7}Ge_{0.3}$ tunnel barrier, a 10-nm Si quantum well(with
phosphorous donor array embedded in the middle), a 7nm undoped
$Si_{0.7}Ge_{0.3}$ tunnel barrier, a 5nm $Si_3N_4$ layer and
lithographically-patterned metallic top gates.   The logic qubit
is represented by the 2-dimensional subspace $|{\bf S},{\bf
S_z}\rangle = |\frac{1}{2},\frac{1}{2}\rangle$ of the three
neighboring donor spins, with logic zero $|0_L\rangle =
|S\rangle|\uparrow\rangle$ and logic one $|1_L\rangle
=\sqrt{2/3}|T_{+}\rangle|\downarrow\rangle-\sqrt{1/3}|T_{0}\rangle|\uparrow\rangle$.
}
\end{figure}
\begin{figure}
\caption{The energies of two lowest states of a single Si:P donor
and their splitting as a function of quantum well width.  For
convenience, the valley splitting is plotted in negative values.}
\end{figure}
\begin{figure}
\caption{ The comparison of the averaged 1D charge distribution of
Si:P ground state in a quantum well with width 6nm and 10nm. (a)
and (b) shows the charge distribution along z and x axis,
respectively. The inset of (b) shows the charge distribution along
y axis.}
\end{figure}
\begin{figure}
\caption{ The valley splitting of (a) Si:P donor ground state and
(b) lowest quantum well state, as a function of quantum well
width. The dotted lines connect points with a spacing of $a_0/2$.
The inset of (a) shows the valley splitting of Si:P ground state
as a function of donor position shift from the quantum well
central plane, for a QW width fixed at 9nm. }
\end{figure}
\begin{figure}
\caption{The gate potential(measured in units of $Ry^*$) profile
at different gate voltages: (a) $V_c = -2.1 V$; (b)$V_c = -0.8 V$;
(c)$V_c = -0.6 V$ and (d)$V_c = -3.2 V$.  $V_g=-1.8$ for all these
four cases.  In each figure, the potential shape is plotted along
x axis, with three different distances(15, 17, 19nm) from the top
gates. The doping plane in our QC device is 17nm below the top
gates.}
\end{figure}
\begin{figure}
\caption{ The exchange coupling of a donor pair as a function of
quantum well width.  The donor separation is fixed at 10
${a_B}^*$. The inset shows the change of exchange coupling as the
QW width is varied in atomic scale, where the calculating points
are spaced by $a_0/4\simeq 1.36\AA$.}
\end{figure}
\begin{figure}
\caption{ (a) The exchange coupling of a donor pair separated by
10 ${a_B}^*$ as a function of central gate voltages. The three
curves correspond to three different values of side gate voltages
$V_g=-1.4, -1.8$ and  $-2.0 V$, respectively. The isolated point
(filled square) on the $V_c=0$ axis corresponds to the exchange
coupling without any gate potential($V_c=V_g=0$). (b) The
dependence of exchange coupling on central gate voltages, with
$V_g$ fixed at -1.8 V. The three curves correspond to three
different donor separations $R=8$, 10, and $12 {a_B}^*$,
respectively. Both (a) and (b) are plotted in logarithmic scale
for the exchange coupling. The quantum well width is fixed at 10
nm.}
\end{figure}
\begin{figure}
\caption{ The averaged 1D two-electron charge distribution of
singlet (solid) and triplet (dotted) states along the inter-donor
axis, for different gate voltages.  The side gate voltage is
$V_g=-1.8 V$ for (b), (c) and (d). The charge distributions for
singlet and triplet almost coincide for weakly coupled donor
pairs, as shown in (a) and (d). The gate-off exchange coupling is
very close to the gate-on case with $V_c=-2.1 V$ and $V_g=-1.8 V$,
while their charge distributions are apparently different, as
shown in (a). It is clearly demonstrated that the gate voltages
can tune the overlap between neighboring donor electron wave
functions and consequently, the exchange coupling.}
\end{figure}
\begin{figure}
\caption{  The exchange coupling as a function of donor position
shift away from the quantum well central plane: (a)gate off; (b)
gate on($V_c=-1.6 V$ and $V_g=-1.8 V$).  The donor pair remains on
the [100] axis, with their separation fixed at 10 ${a_B}^*$.  The
inset of (a) plots the overlap of single-particle wave
functions(for singlet state) as a function of donor position
shift, which shows the same pattern as the exchange coupling.
 The dotted line in (b) is an exponential fitting curve,
illustrating the significant effect of gate potential on the
exchange coupling for donors at different distances from top
gates.}
\end{figure}
\begin{figure}
\caption{ The exchange coupling(plotted in logarithmic scale) as a
function of donor separation.  The donor pair remains on the [100]
axis, with their separation varied from 70 to 90, in units of half
lattice constant, $a_0/2$. The dotted line is an exponential
fitting curve, appearing as a straight line in the
logarithmic-scale plotting .}
\end{figure}

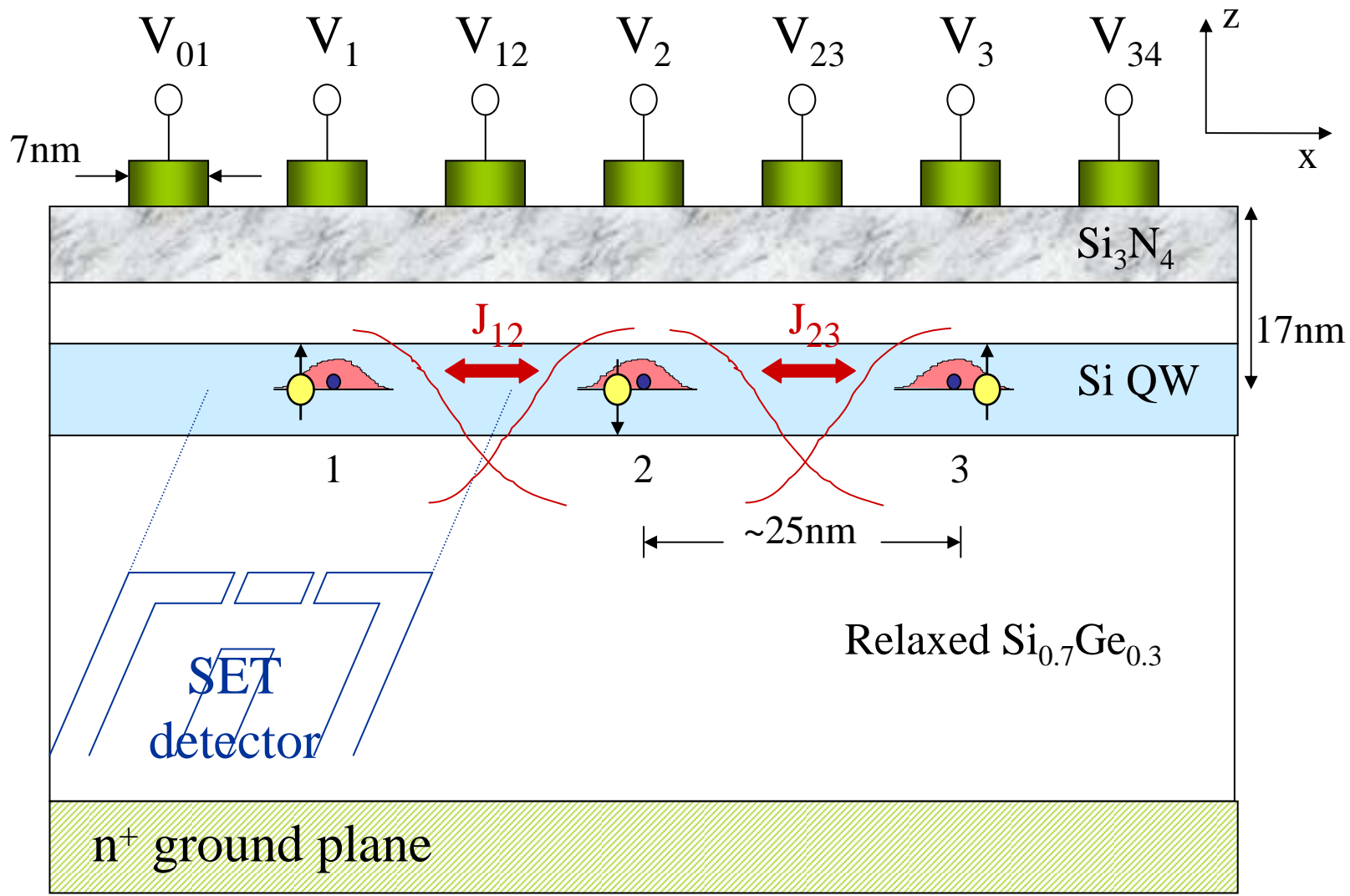
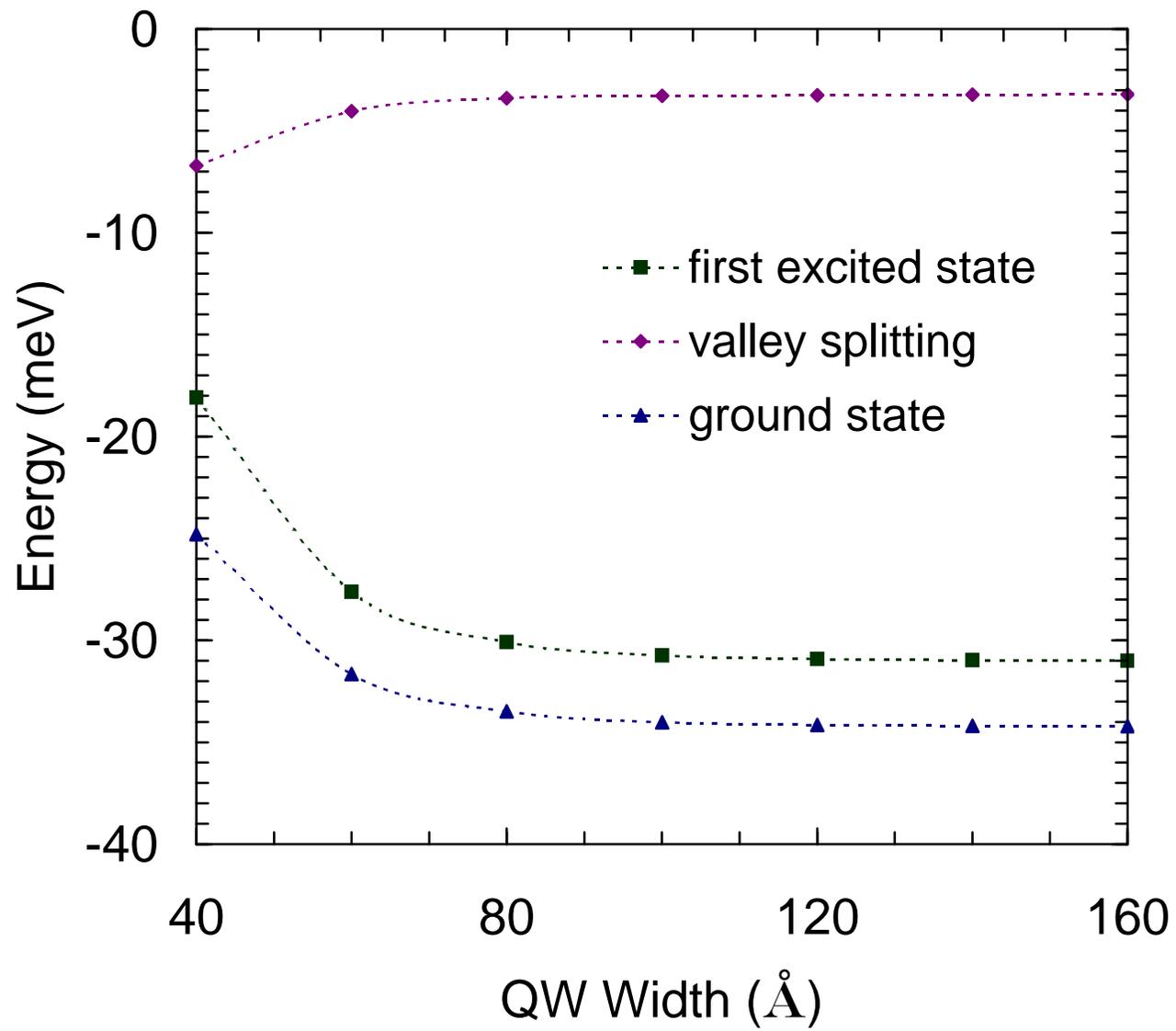
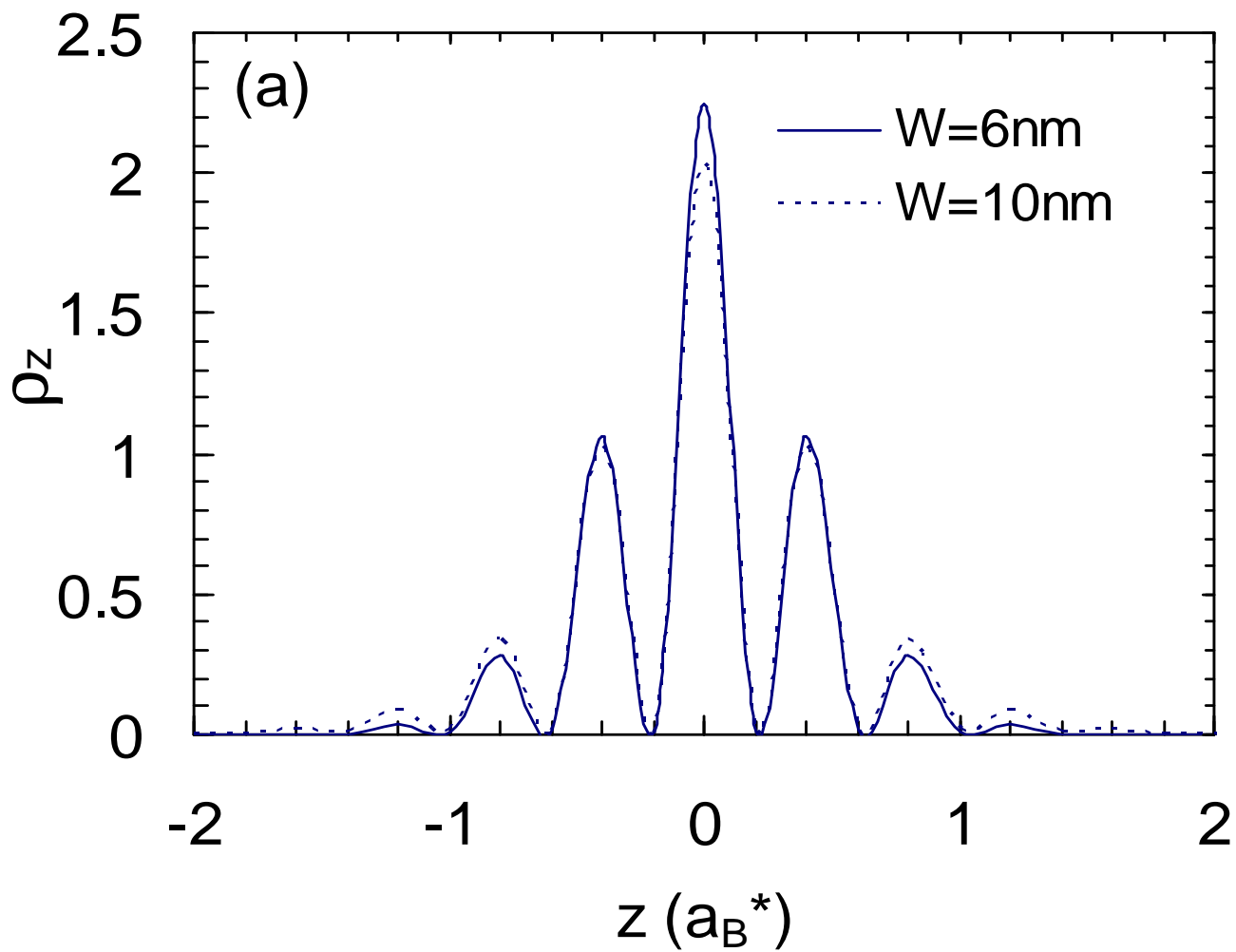
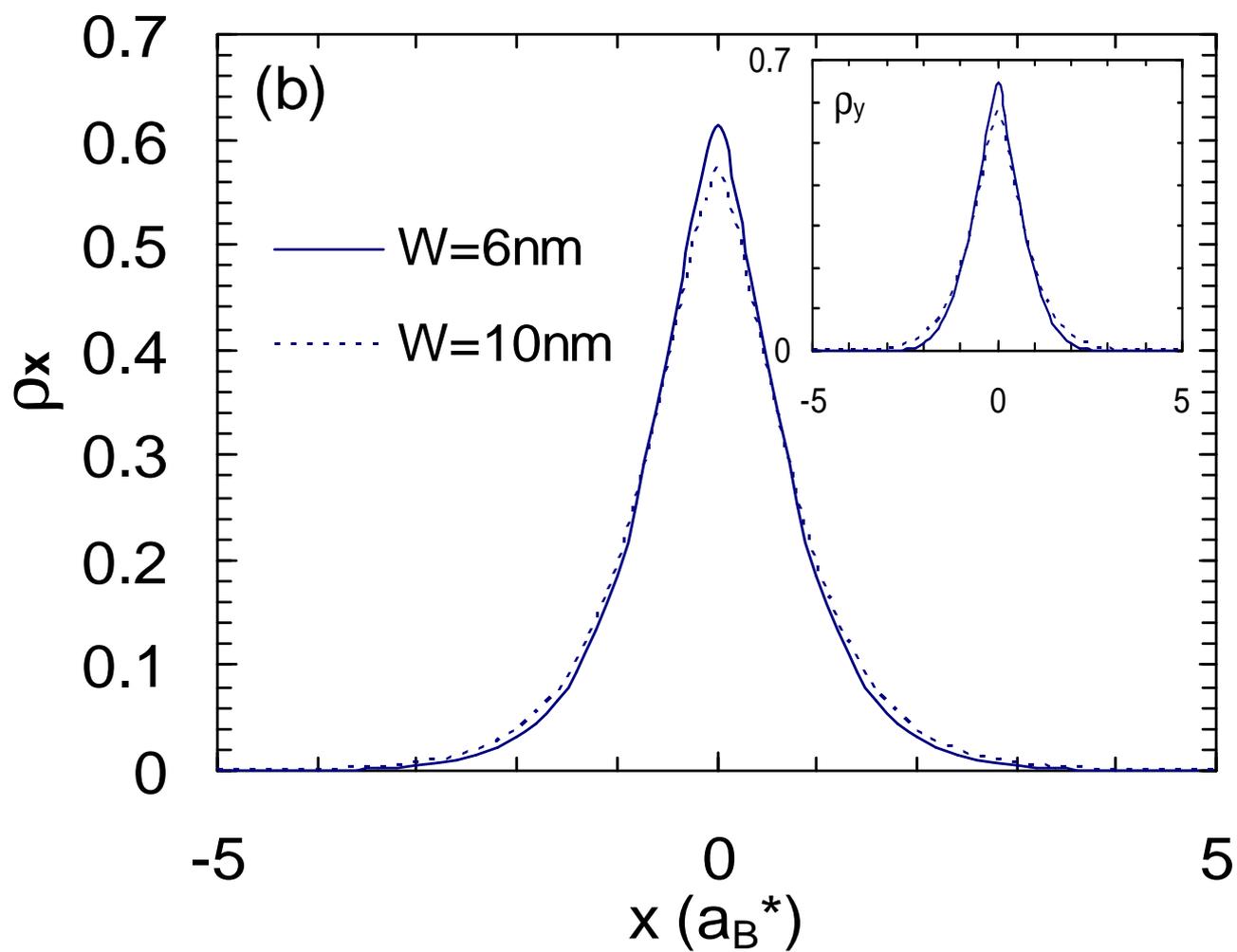
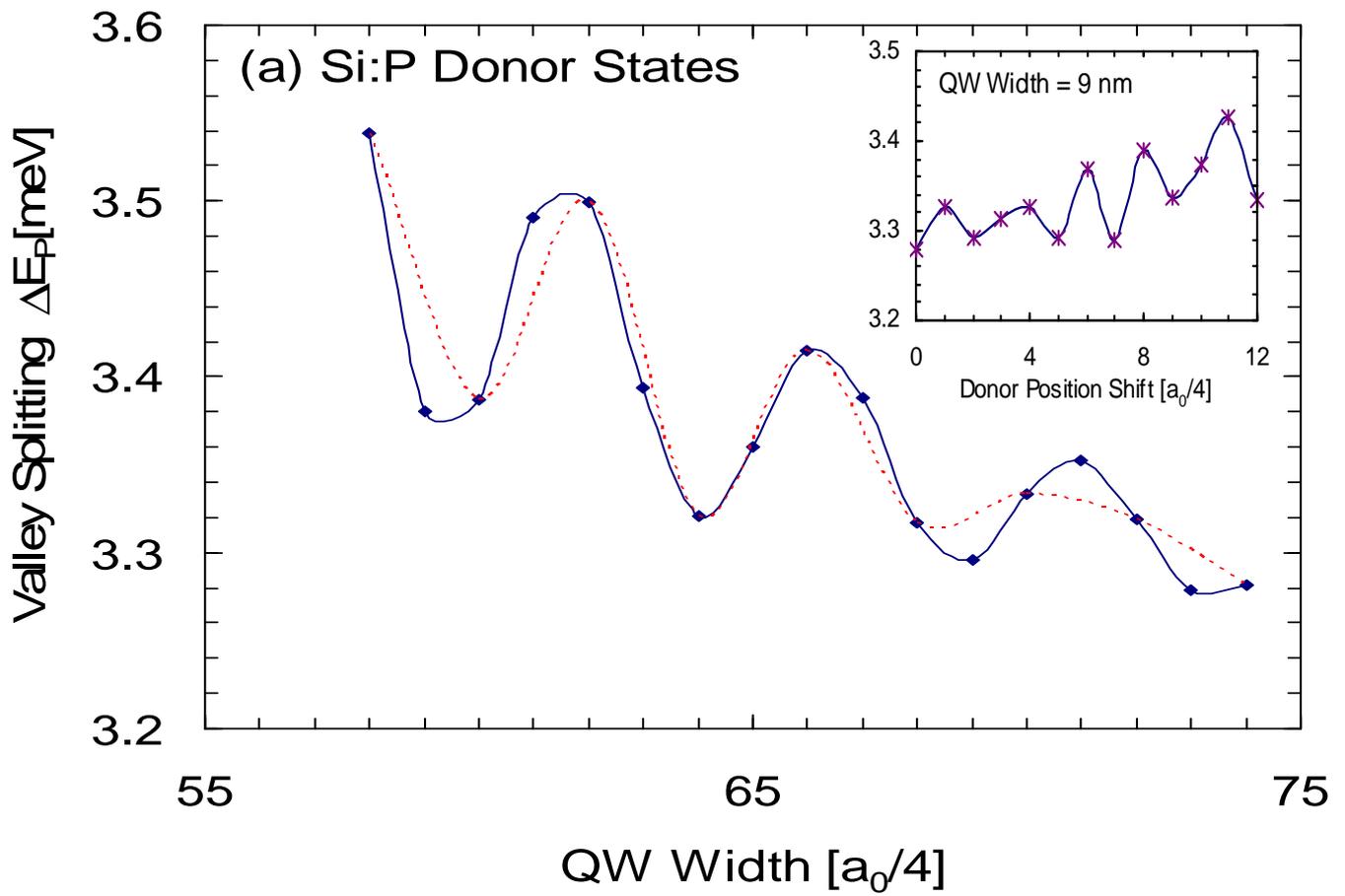
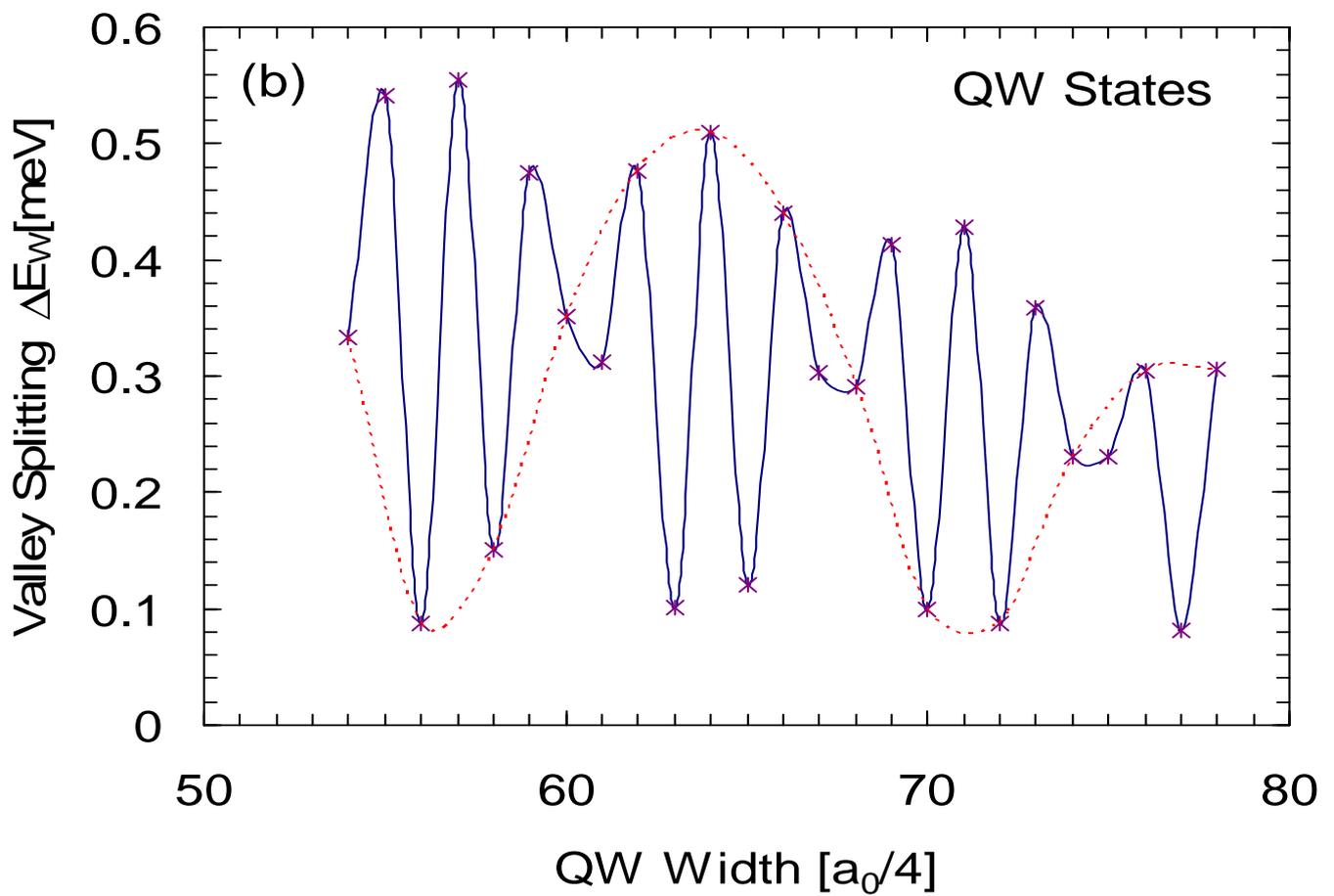
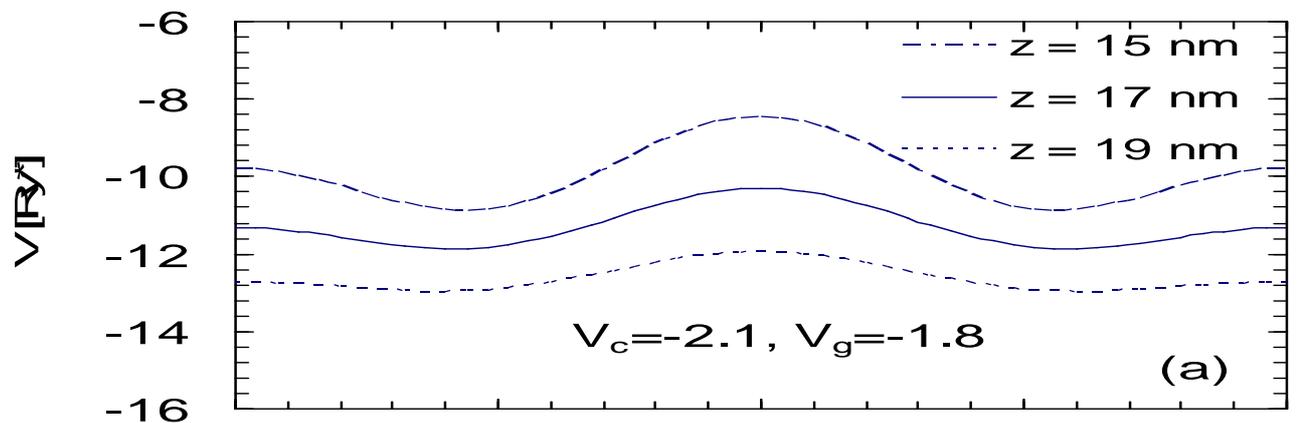
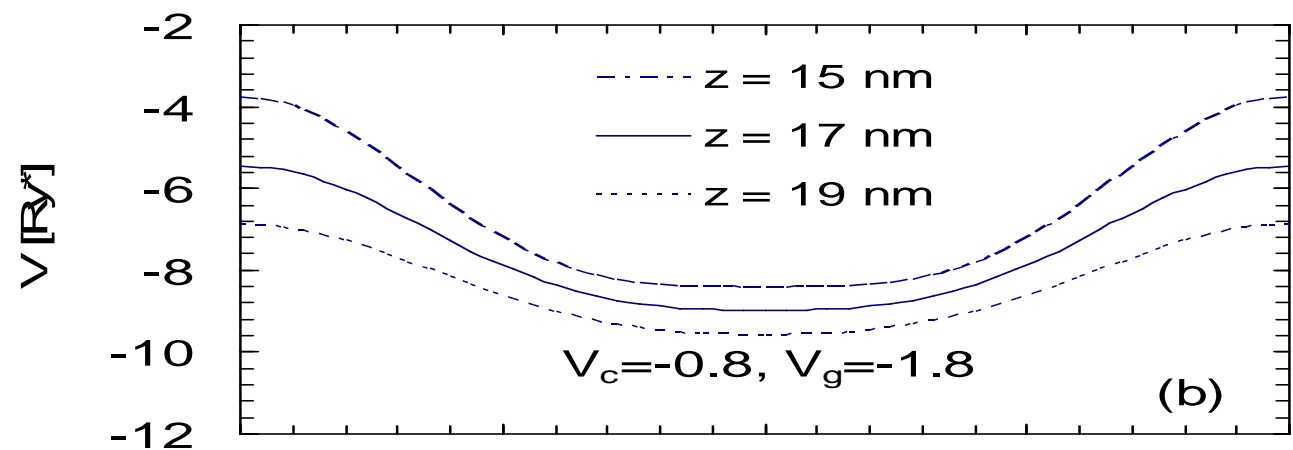
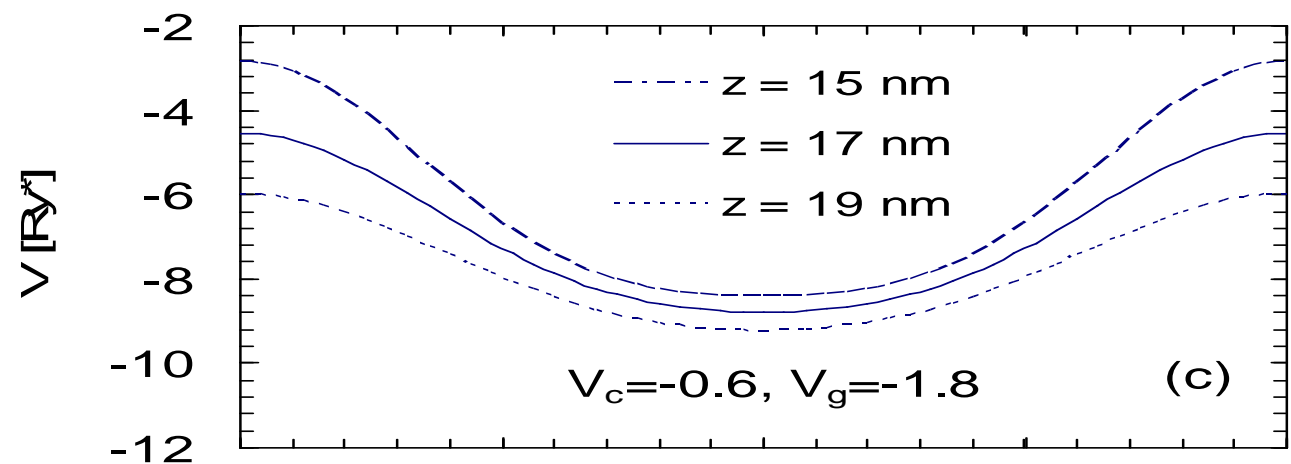
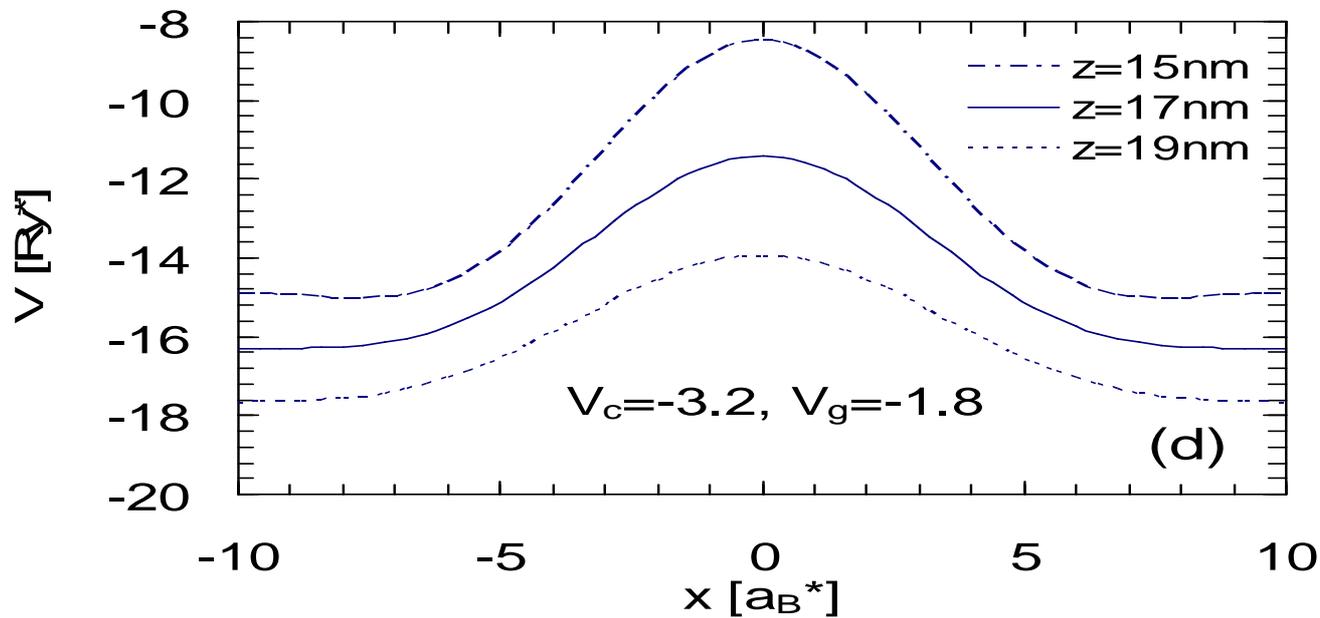
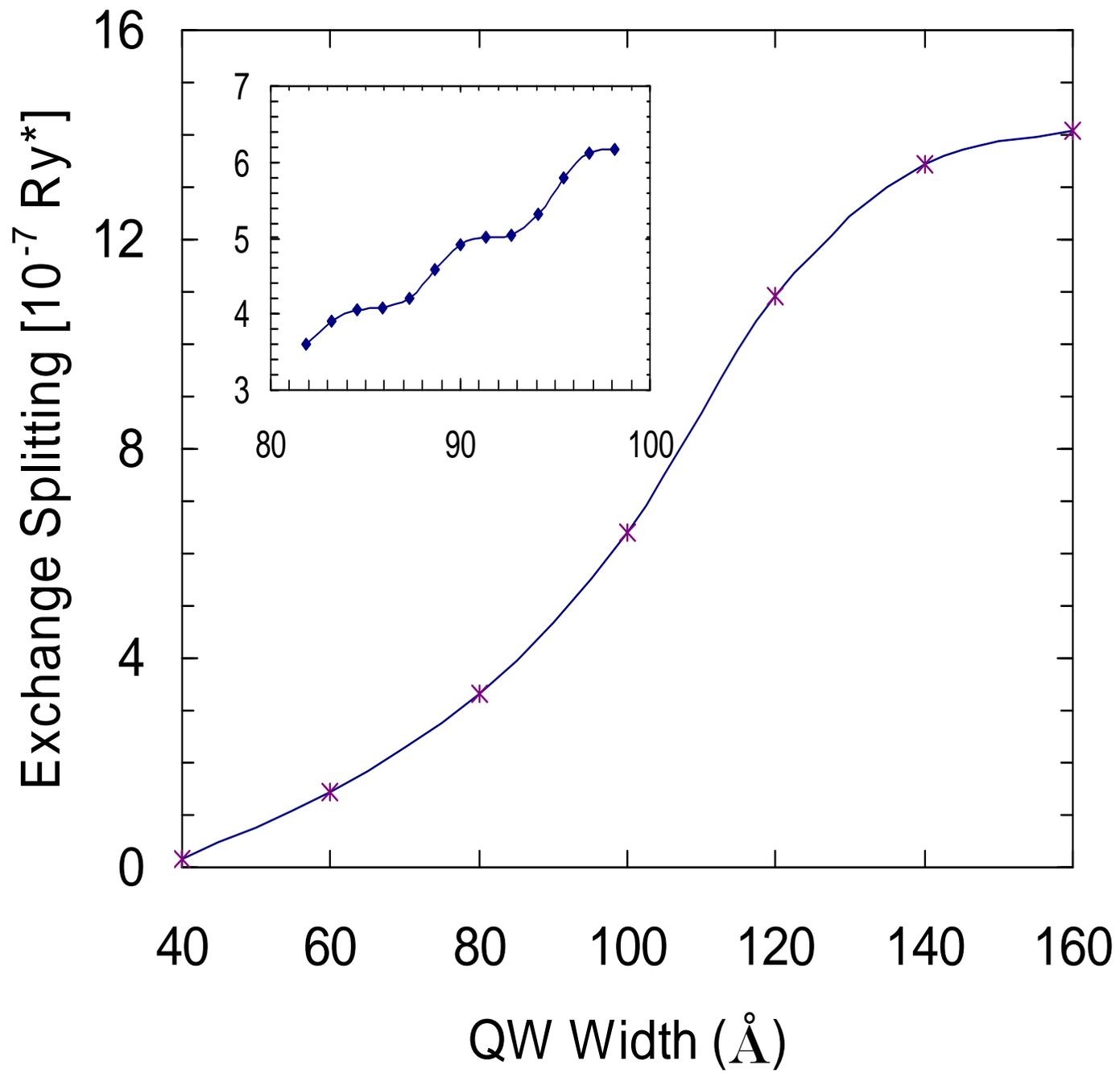
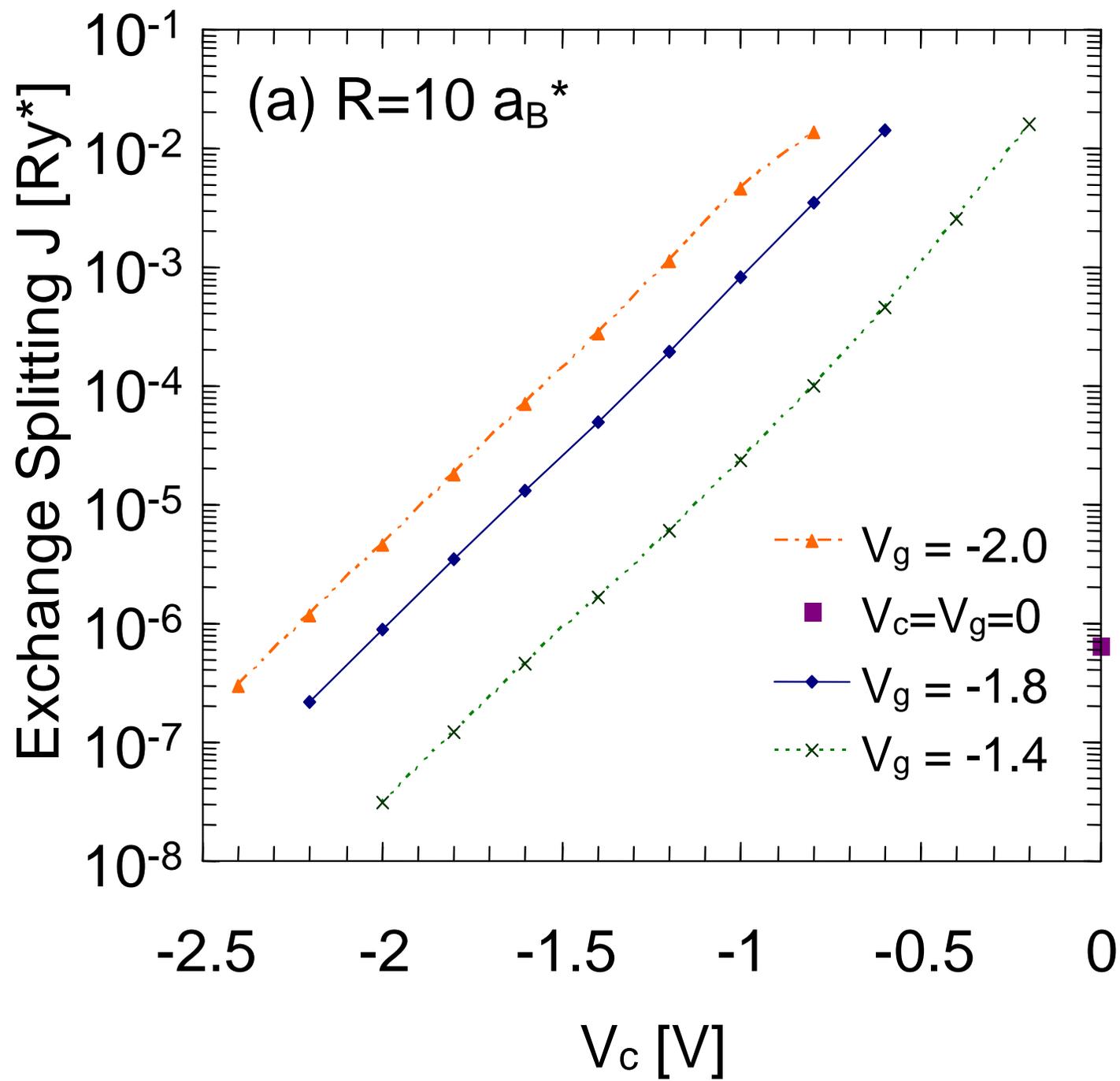
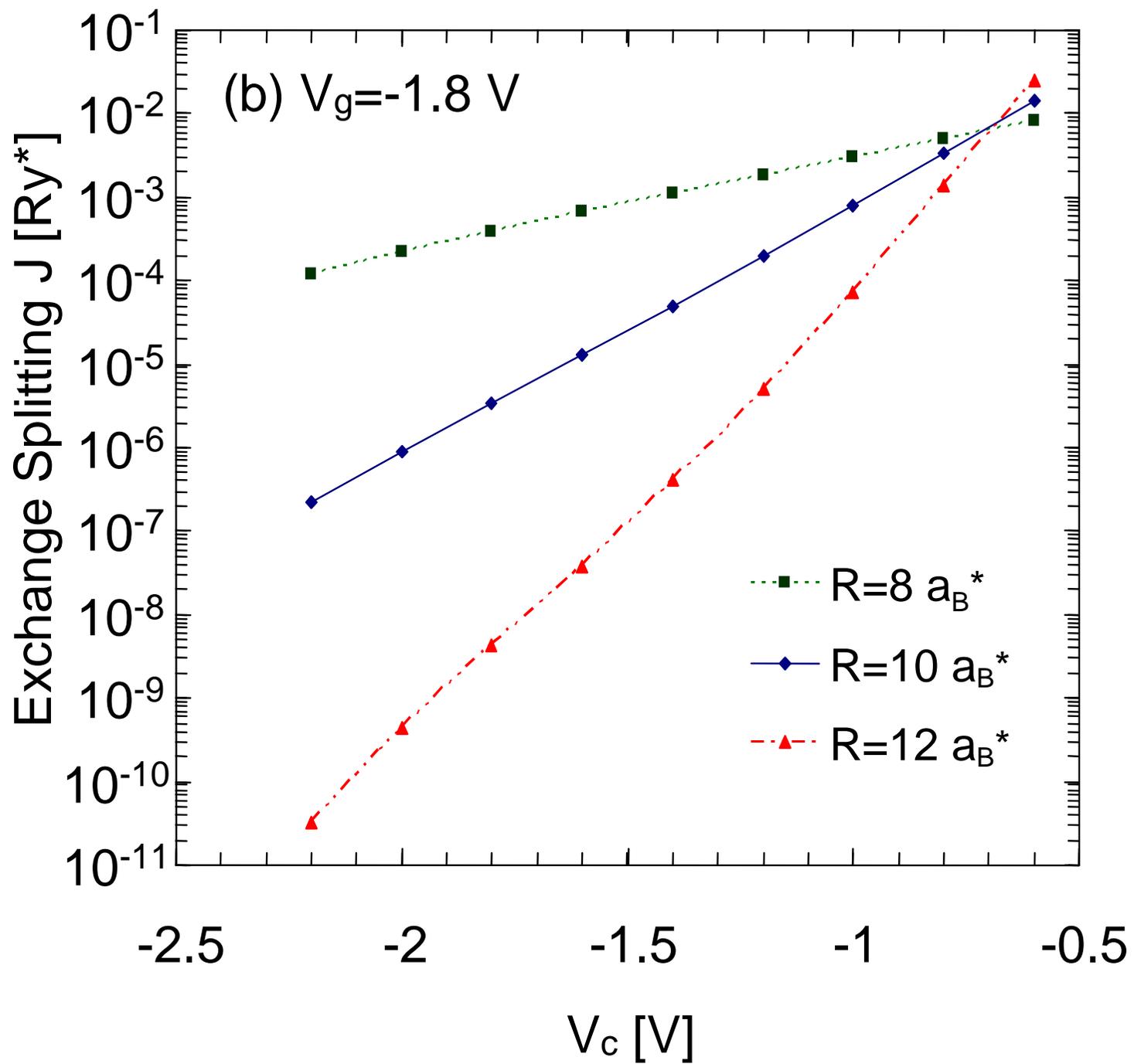
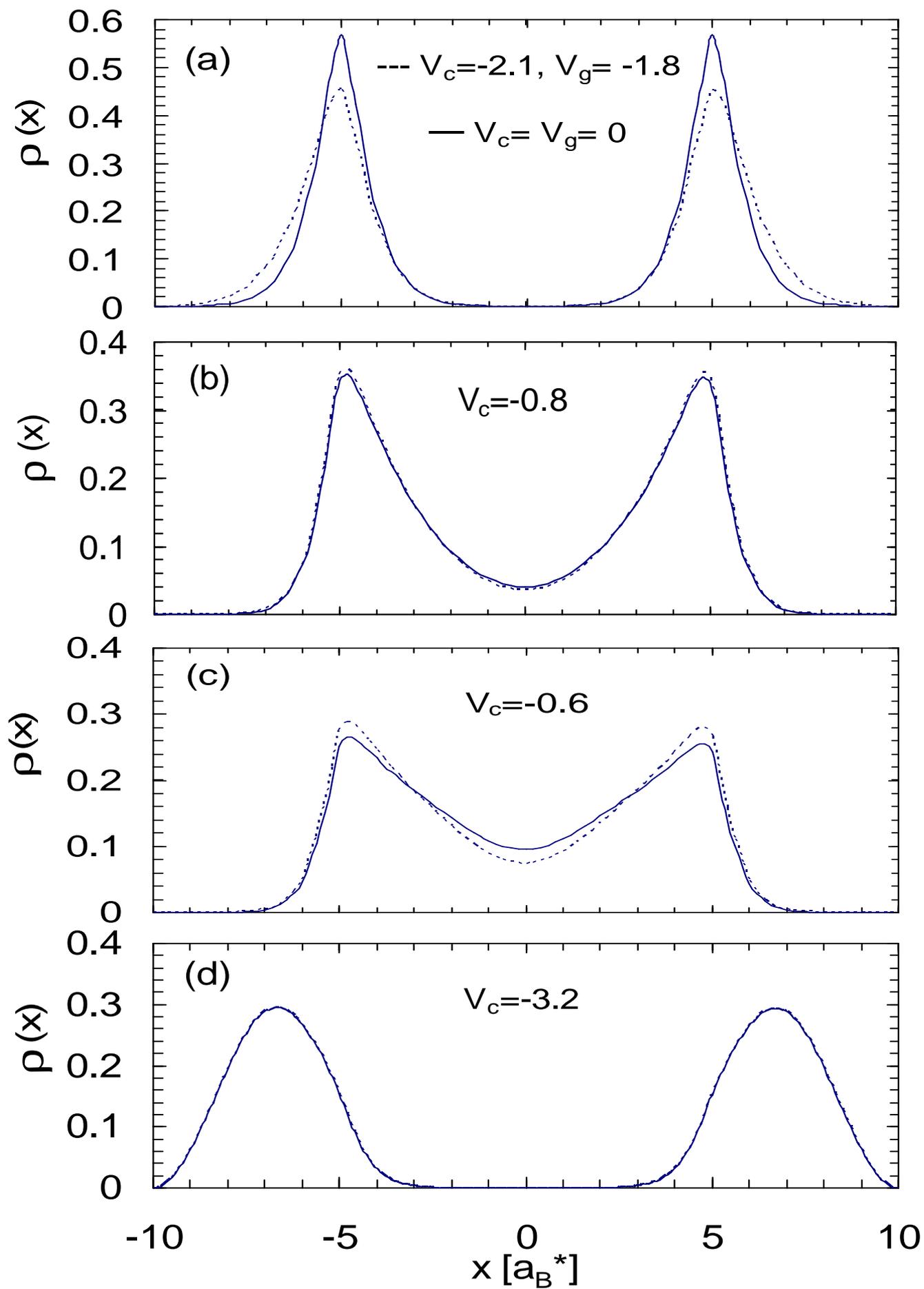
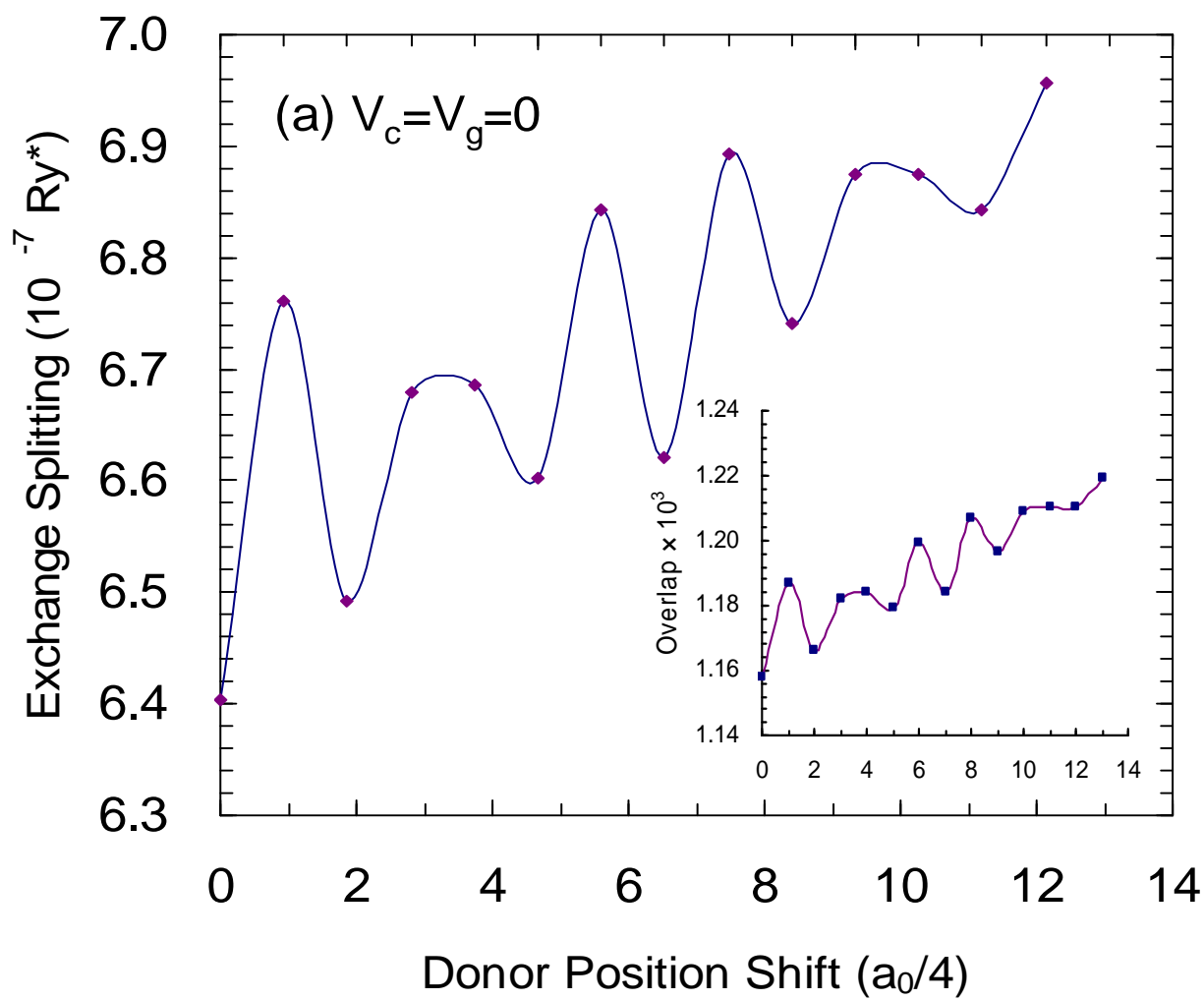
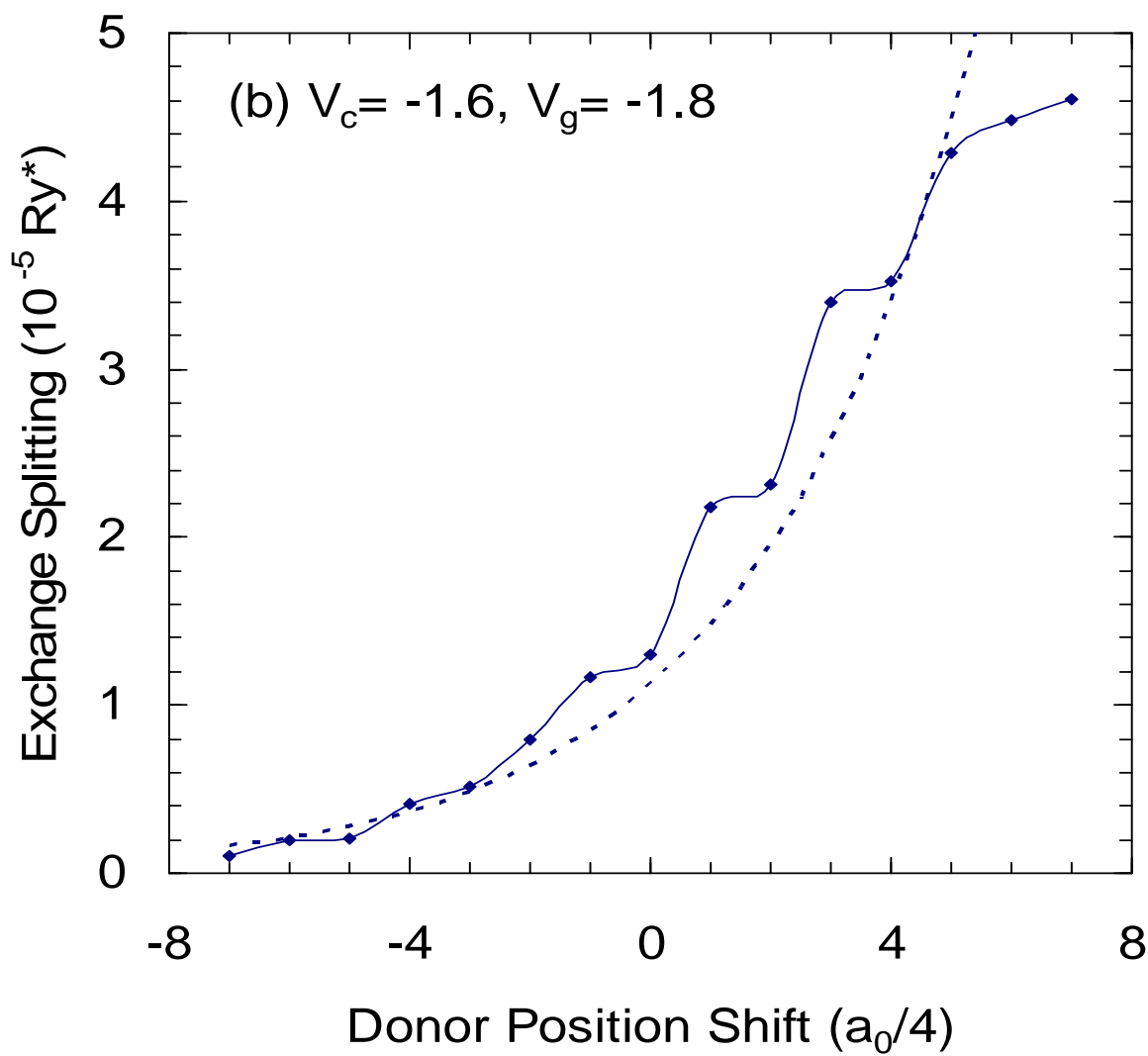
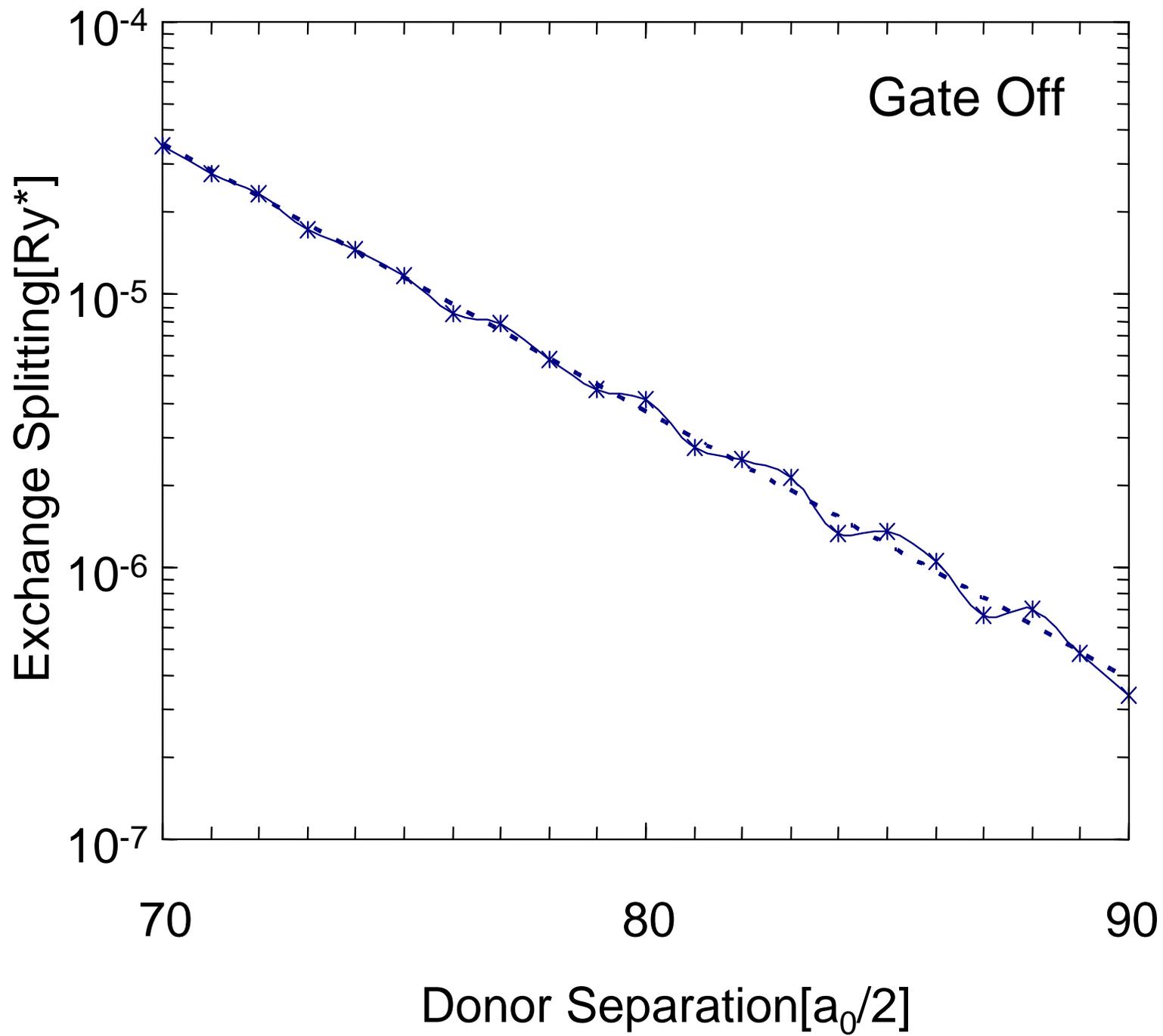
\end{document}